\documentclass[twocolumn,aps,showpacs,showkeys,prb
   ,superscriptaddress,reprint]{revtex4-1}
\usepackage{natbib}
\usepackage{url}
\usepackage{bm}
\usepackage{textcase}
\usepackage{color}
\usepackage{graphicx}
\usepackage{amsmath}
\usepackage{ulem}
\usepackage{url}
\usepackage{textcase}
\usepackage{graphicx}
\usepackage[colorlinks=true,citecolor=blue]{hyperref} %specify this as last package
\hypersetup{pdftitle={Beam matching: a method to study transport across multiple
  interfaces}}
\hypersetup{pdfauthor={Debanjan Basu, Peter E. Bloechl}}
\hypersetup{pdfdisplaydoctitle}

\newcommand{\mat}[1]{{\boldsymbol #1}}
\newcommand{\e}[1]{{\rm e}^{#1}}

\renewcommand{\Im}{{\rm Im}}
\renewcommand{\Re}{{\rm Re}}

\newcommand{\arccosh}{{\rm arccosh}}
\begin{document}
\title{Beam matching: a method to study phonon transport
  through interfaces and multilayer structures }
\author{Debanjan Basu}

\affiliation{Institute for Theoretical Physics, Clausthal University
  of Technology, Leibnizstr. 10, 38678 Clausthal-Zellerfeld, Germany}

\author{Peter E. Bl\"ochl}

\affiliation{Institute for Theoretical Physics, Clausthal University
  of Technology, Leibnizstr. 10, 38678 Clausthal-Zellerfeld, Germany}

\affiliation{Institute for Materials Physics,
  Georg-August-Universit\"at G\"ottingen, Friedrich-Hund-Platz 1,
  37077 G\"ottingen, Germany}

\date {\today}
 
\begin{abstract}
 Structuring materials is one mechanism to influence the
  thermal conductivity and thus thermoelectric efficiency. In order to
  investigate the scattering of phonons in multilayer structures we
  developed a beam matching technique, which is based on the concept
  of individual phonons and their scattering at interfaces. One of the
  major goals is to efficiently determine the complex
  band structure of the bulk materials.  The complex band structure is
  determined using selected k-points on a triangulated grid in the
  complex plane of wave vectors perpendicular to the interface.  
  Matching the phonon modes at an interface is translated to a
  singular value problem. Its null-vectors provide the coupling
  coefficients of the phonon modes across the interface. Besides
  giving explicit access to the modes as they
  scatter at the interface, the technique provides the transfer
  matrices, that provide the transmission coefficient of any
  multilayer structure.  The transmission coefficient, in
  turn, yields the phononic thermal conductance for coherent
  transport. The knowledge of the matched phonons forms the basis of
  investigating incoherent transport under the influence of
  phonon-phonon or impurity scattering.
\end{abstract}
\maketitle

%=====================================================================
\section{Introduction}
%=====================================================================
Control of thermal conductivity is essential for the development of
thermoelectric devices. Thermoelectric efficiency depends on the
presence of a large electronic conductivity and a small thermal
conductivity. The so-called figure of merit $ZT$,
which describes how efficiently a material can produce thermoelectric
power, is given by
\begin{eqnarray}
ZT=\frac{\sigma S^2 T}{\kappa}\;,
\end{eqnarray}
where $\sigma$ is the electric conductivity, $S$ is the Seebeck
coefficient, $T$ is the temperature and $\kappa$ is the thermal
conductivity. Besides electrons, phonons are the dominant
carriers for thermal conductivity.  Because the thermal
conductivity due to electrons is closely linked to the electric
conductivity via the Wiedemann-Franz law, the decrease of phononic
thermal conductivity is one handle to improve the efficiency of
thermoelectric devices.

Structuring materials is one possibility to influence the phononic
thermal conductivity. The basic structuring element is an interface,
which acts like a filter for the impinging phonons. Phonons on one
side of the interface will be partially reflected by the interface and
partially transmitted across the interface. In certain frequency
regions, one of the materials may not support propagating phonons,
which completely blocks out the corresponding phonon current. By
alternating soft and hard materials, for example, the thermal
resistance can be increased above the mean resistance of both
materials. The thermal resistance of an interface, called Kapitza
resistance\cite{kapitza41_jpussr4_181}, produces a finite temperature
step across the interface in the presence of a thermal current.

More complex structures involve several interfaces. Multiple
reflections between these interfaces enhance or suppress the thermal
current due to constructive or destructive interference
respectively. These effects depend crucially on the relative phases of
individual phonons.  Furthermore, the local behavior in structures
with small spatial dimensions differs from the near-equilibrium
description employed on macroscopic length scales.

This work addresses the transport of phonons across single and multiple
interfaces.  The tools developed can, however, be extended to the
transport of other quasiparticles across single or multiple
interfaces.  The description accounts for evanescent waves, which
mediate phonon tunneling through thin layers of a material, and which
can also transport heat across the interface, if additional scattering
processes convert the energy into propagating modes. Evanescent waves
can combine to form true interface states, that exist only in the
proximity of the interface. The methods preserve the phase
information, which allows the study of interference effects at
multiple interfaces.

The focus on the propagating and evanescent waves as the
  elementary entities sets this method apart from the atomistic Green's
  function
  method\cite{mingo03_prb68_245406,sadasivam14_annrevheattransfer17_89,wang08_epjb62_381,zhang07_numerheattransferb51_333,ong15_prb91_174302},
  an elegant and widely used method for phonon transport problems.

The relation between the atomistic Green's function method and
  our mode-based method is analogous to that of Green's-function based
  and wave-function based approaches in electronic structure
  theory. While each method has its own virtues, they can be
  transformed into each other. For example, the mode coupling
  coefficients can be extracted from the Greens
  function\cite{ong15_prb91_174302}.

The advantage of the Green's function method over the present
  method is the ease with which ensemble averages, for example over
  scatterers, are performed.  The mode matching described here, in
  contrast, makes the contact to the individual modes more direct.

The present method also differs from many previous approaches
  \cite{chang82_prb25_605,chang82_prb25_3975,yip84_prb30_7037} in that
  it first evaluates the complex band structure, where it exploits the
  full symmetry of the bulk materials.  These waves are then matched
  at an interface, which has a reduced symmetry.

In the remainder of this section, we briefly review the theoretical
basis of thermal transport via phonons in the incoherent and the
coherent limits. Thus we are revisiting the Boltzmann equation on the
one hand and the Landauer formula on the other. Both take the complex
band structure of the individual materials and the transmission
coefficients as the basic input, which are the topic of the present
paper. In section~\ref{sec:eqm}, we relate this description to the
motion of individual atoms. These sections serve to introduce our
notation. Readers specifically interested in the methodology
  of calculating the complex band structure and beam matching may skip
  the introductory sections and return to them whenever the notation
  needs clarification. In section~\ref{sec:complexbandsstructure} we
introduce the concept of a complex band structure. The methods have
been analyzed using model systems described in
section~\ref{sec:modelsystes}.  The method for the numerical
determination of the complex band structure is presented in
section~\ref{sec:numericscomplexbands}.  In
section~\ref{sec:beammatching}, we describe how the individual
solutions are combined at an interface, and how the resulting transfer
matrices can be used to determine the transmission through complex
multilayer structures. In section~\ref{sec:fromtransfertoconduct} we
describe how to extract the conductance and explore their spectral
distribution for a model interface.

%=====================================================================
\subsection{Thermal conductivity from the phonon-Boltzmann equation}
\label{sec:boltzman}
%=====================================================================
The thermal conductivity tensor $\mat{\kappa}$ relates the
heat-current density $\vec{j}_Q$ to the temperature gradient
$\vec{\nabla}T$,
\begin{eqnarray}
\vec{j}_Q=\sum_{\lambda} n_\lambda \epsilon_\lambda\vec{v}_\lambda
= -\mat{\kappa}\vec{\nabla}T\;,
\label{eq:thermalcurrent}
\end{eqnarray}
% definition crosscheck: https://en.wikipedia.org/wiki/Thermal_conduction
where $n_\lambda$ is the local phonon density in the mode specified by
$\lambda$.  $\lambda=(\vec{k},\sigma)$ combines the quantum numbers of
a phonon, namely its wave vector $\vec{k}$ and band index $\sigma$.
The energy of a phonon
$\epsilon_\sigma(\vec{k})=\hbar\omega_\sigma(\vec{k})$ and its
velocity
$\vec{v}_\sigma(\vec{k})=\vec{\nabla}_k\omega_\sigma(\vec{k})$ are
obtained from the dispersion relation, that is, the phonon
band structure $\omega_\sigma(\vec{k})$.

The phonon occupations in a temperature gradient can be extracted from
the phonon-Boltzmann
equation\cite{peierls29_ap3_1055,klemens58_ssp7_1}, which describes
the dynamics of the phonon densities $n_\lambda(\vec{r},t)$ in space
and time. After linearizing the scattering term in the deviation of
the phonon densities from their equilibrium values, the
resulting linearized phonon-Boltzmann equation is
\begin{eqnarray}
\frac{\partial n_\lambda}{\partial t}&+&\vec{v}_\lambda\vec{\nabla} n_\lambda
=-\sum_{\lambda'} B_{\lambda,\lambda'}\Bigl(n_{\lambda'}-n^{eq}_{\lambda'}(T)\Bigr)
\;.
\label{eq:phononboltzmann}
%% \nonumber\\
%% \stackrel{dn_\lambda/dt=0}{\Rightarrow}
%% n_\lambda&=&n^{eq}_\lambda(T)
%% +\sum_{\lambda'}B^{-1}_{\lambda,\lambda'}
%% \vec{v}_{\lambda'}\vec{\nabla} n_{\lambda'}
\end{eqnarray}
The first order of the scattering term in deviations from thermal
equilibrium defines $\mat{B}$ as
\begin{eqnarray}
B_{\lambda,\lambda'}=-\left.\frac{\partial}{\partial{n}_{\lambda'}}\right|_{n^{eq}(T)}
\left(\frac{dn_\lambda}{dt}\right)_{\text{scatt}}
\;.
\end{eqnarray}
The equilibrium phonon density
$n^{eq}_\lambda=n^0_\lambda\left[\e{\beta\epsilon_\lambda}-1\right]^{-1}$
is given by the Bose distribution and the density $n^0_\lambda$ of a
single phonon in mode $\lambda$.  The latter is described in the
discussion following Eq.~\ref{eq:phonondensity} below. As
  usual, $\beta=1/(k_BT)$.

For stationary distributions $n_\lambda$, the thermal current can be
extracted from Eqs.~\ref{eq:thermalcurrent} and
\ref{eq:phononboltzmann} 
\begin{eqnarray}
\vec{j}_Q=\sum_{\lambda} n_\lambda \epsilon_\lambda\vec{v}_\lambda
=-\sum_{\lambda,\lambda'}
\epsilon_\lambda\vec{v}_\lambda
B^{-1}_{\lambda,\lambda'}
\Bigl(\vec{v}_{\lambda'}\vec{\nabla} n_{\lambda'}\Bigr)\;.
\end{eqnarray}
The equilibrium distribution does not show up, because it does not
contribute any net currents.  With
$\vec{\nabla}n_\lambda=\frac{dn^{eq}_\lambda}{dT}\vec{\nabla}T$, the
thermal conductivity tensor is obtained, 
\begin{eqnarray}
\mat{\kappa}&=&\sum_{\lambda,\lambda'}
B^{-1}_{\lambda,\lambda'}
\epsilon_\lambda\frac{dn^{eq}_{\lambda'}}{dT}
\vec{v}_\lambda\otimes\vec{v}_{\lambda'}
\;,
\end{eqnarray}
where $\otimes$ denotes the outer product
$\left(\vec{a}\otimes\vec{b}\right)_{i,j}=a_ib_j$.

\begin{figure}[h!]
\begin{center}
\includegraphics[width=0.8\linewidth]{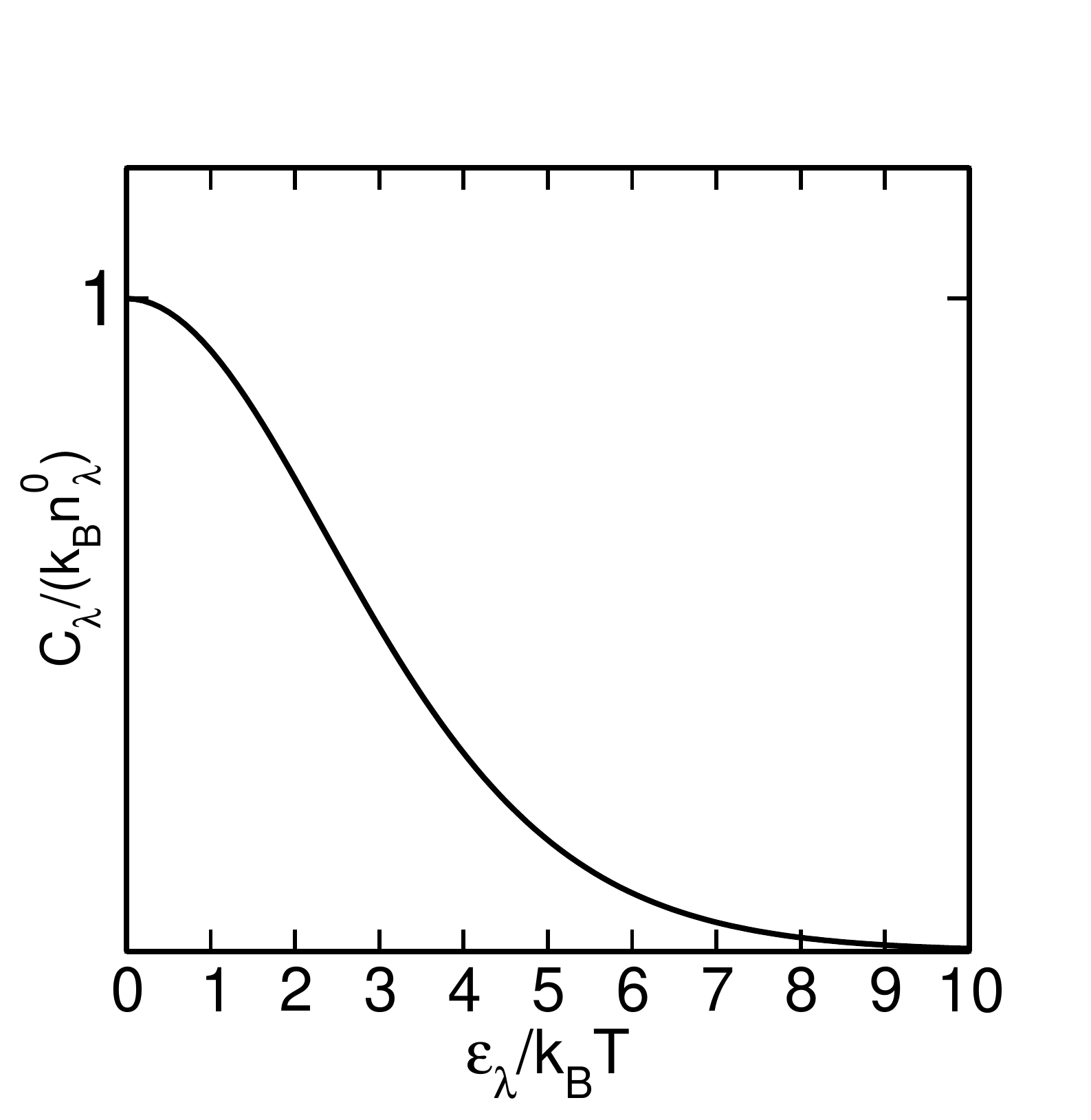}
\end{center}
\caption{\label{fig:heatcapacity}Heat capacity $C_\lambda$ per volume
  of a phonon mode as function of the phonon energy
  $\epsilon_\lambda$.}
\end{figure}

Considering only the diagonal elements of $\mat{B}^{-1}$ in the phonon
modes, one obtains the well known
relation\cite{klemens58_ssp7_1,turney09_prb79_64301} for
  thermal conductivities
\begin{eqnarray}
\mat{\kappa}=\sum_\lambda C_\lambda\tau_\lambda 
\vec{v}_\lambda\otimes\vec{v}_\lambda\;,
\label{eq:kappafromphononboltzmann}
\end{eqnarray}
where $\tau_\lambda=(\mat{B}^{-1})_{\lambda,\lambda}$ can be
identified with the relaxation time, respectively the phonon lifetime.
$C_\lambda=\epsilon_\lambda dn^{eq}_\lambda(T)/dT$ is the contribution
of a phonon mode to the heat capacity per unit volume.  A graph of
$C_\lambda$ is shown in Fig.~\ref{fig:heatcapacity}. Its main effect
is to cut off the high-frequency phonons at low temperatures,
i.e. below the Debye temperature.

Among the main bulk scattering processes are impurity scattering,
electron-phonon scattering and the temperature dependent phonon-phonon
scattering\cite{omini95_physicab212_101,broido05_prb72_14308,chernatynskiy10_prb82_134301}.

One of the central approximations of the Boltzmann equation is that
the phase information of phonons is lost between two scattering
events. The phase information is responsible for quantum mechanical
interference effects. This approximation implies that the typical
scattering length needs to be larger than the coherence length.

The importance of this limitation is evident from the two possible
descriptions of a multilayer. One uses the modes of the
multilayer. The other uses the modes of the individual layers and the
transmission coefficients across the interfaces. The first description
takes coherent multiple scattering between interfaces fully into
account, while the second suppresses this effect by introducing a
phase average after each interface
traversal. Applying the Boltzmann equation to
multilayer systems is thus not unique. It needs to be done with
attention to the relevant decoherence mechanisms and dissipation
channels.

%=====================================================================
\subsection{Thermal conductance from ballistic transport}
\label{sec:landauer}
%=====================================================================
The assumptions of the Boltzmann equation require that the length
scale of structured materials is larger than the coherence length of
the phonons. For materials with structures on the nano-scale the
underlying assumptions of the Boltzmann equation are often violated.

The opposite extreme to the incoherent transport described by the
Boltzmann equation is ballistic transport. In coherent or ballistic
transport, multiple scattering is considered, i.e. scattering events
cannot be disentangled because they occur within the coherence length
of the scattered particle.

The heat current $I_Q$ through the material structure is related to
the temperature difference $\Delta T$ at the contacts by the
conductance $G$.
\begin{eqnarray}
I_Q=G\Delta T
\label{eq:heatcurrentballistic}
\end{eqnarray}
The thermal conductance of a homogeneous block of material can be
expressed by its thermal conductivity $\mat{\kappa}$ as
$G=\frac{A}{\ell}(\vec{e}\mat{\kappa}\vec{e})$, where $A$ is the
cross section of the material, $\ell$ is its length, and $\vec{e}$ is
the normal vector of the cross-section plane.

The ballistic heat current is obtained 
%
% with the basic assumption of
% the Landauer formula\cite{landauer87_zpb68_217}, namely 
%
assuming that the density $n_\lambda$ of
incoming phonons in each contact is identical to the equilibrium
density $n_\lambda^{eq}(T)$ at the temperature of that contact.  The
outgoing phonon fluxes are related to the incoming phonon fluxes by
the reflection and transmission coefficients: the transmission
coefficient $\mathcal{T}_{\lambda,\lambda'}$ is the ratio between the
flux of transmitted phonons in mode $\lambda'$ and the flux of
incoming phonons in mode $\lambda$. Similarly, the reflection
coefficient $\mathcal{R}_{\lambda,\lambda'}$ is the ratio between the
flux of reflected phonons in mode $\lambda'$ and the flux of incoming
phonons in mode $\lambda$.  The reflection and transmission
coefficients account for all multiple scattering events inside the
material.

The ballistic heat current from contact $A$ to contact $B$
is
\begin{eqnarray}
I_Q
=A_A
\hspace{-0.3cm}\sum_{\lambda\in A,\lambda'\in B}\hspace{-0.1cm}&&
\theta(\vec{e}\vec{v}_\lambda)n^{eq}_\lambda(T_A)
(\vec{e}\vec{v}_\lambda)\epsilon_\lambda 
\mathcal{T}^{A\rightarrow B}_{\lambda,\lambda'}
\nonumber\\
-
A_B\hspace{-0.3cm}\sum_{\lambda\in A,\lambda'\in B}\hspace{-0.1cm}&&
\theta(\vec{e'}\vec{v}_{\lambda'})n^{eq}_{\lambda'}(T_B)
(\vec{e'}\vec{v}_{\lambda'})\epsilon_{\lambda'} 
\mathcal{T}^{B\rightarrow A}_{\lambda',\lambda}\;,
\end{eqnarray}
where $\theta(x)$ is the Heaviside step function, which is unity for
positive arguments and which vanishes for negative arguments. It
selects the phonons in the incoming direction.  $A_A$ is the cross
section of contact $A$ and $\vec{e}$ is its plane normal pointing
inward. Similarly, $A_B$ is the cross section of material $B$ with the
plane normal $\vec{e'}$ pointing inward. We exploited that reflection
and transmission coefficients add up to one, i.e.
\begin{eqnarray}
\sum_{\lambda'\in A}\mathcal{R}^{A\rightarrow{A}}_{\lambda,\lambda'}
+\sum_{\lambda'\in
  B}\mathcal{T}^{A\rightarrow{B}}_{\lambda,\lambda'}=1
\;,
\end{eqnarray}
in order to relate the unidirectional flux to the transmitted fluxes.

For equal temperatures, i.e. $T_A=T_B$, the thermal current vanishes,
which relates the right-going flux to the left-going flux at the same
temperature. This notion allows to simplify the expression for the
heat current by replacing the difference of left- and right-going
fluxes by the differences of two right-going fluxes at two different
temperatures.
\begin{eqnarray}
I_Q
=A_A\sum_{\lambda\in A}
&&\Bigl(n^{eq}_\lambda(T_A)-n^{eq}_\lambda(T_B)\Bigr)
\epsilon_\lambda 
\nonumber\\
&&\times
\sum_{\lambda'\in B}\theta(\vec{e}\vec{v}_\lambda)(\vec{e}\vec{v}_\lambda)
\mathcal{T}^{A\rightarrow B}_{\lambda,\lambda'}
\label{eq:heatcurrentballisticfinal}
\end{eqnarray}

For small temperature differences $T_A-T_B$, the equilibrium
occupations can be linearized in temperature so that we obtain the
thermal conductance $G$ from Eqs.~\ref{eq:heatcurrentballistic} and
\ref{eq:heatcurrentballisticfinal} as
\begin{eqnarray}
G=A_A\sum_{\lambda\in A} C_\lambda\;
\sum_{\lambda'\in B}
\theta(\vec{e}\vec{v}_\lambda)
\;(\vec{e}\vec{v}_\lambda)
\mathcal{T}^{A\rightarrow B}_{\lambda,\lambda'}
\;.
\label{eq:conductancelandauer}
\end{eqnarray}
$C_\lambda=\epsilon_\lambda\frac{\partial}{\partial{T}}
n^{eq}_\lambda$ also occurred in the expression
Eq.~\ref{eq:kappafromphononboltzmann} for the thermal conductivity. It
is the contribution of the phonon mode $\lambda$ to the heat capacity
per volume. It is evaluated at the mean temperature
$T=\frac{1}{2}(T_A+T_B)$ of the two contacts.

The basic quantities determining the thermal current due to phonons
are (1) the phonon dispersion relation $\omega_\sigma(\vec{k})$, (2)
the equilibrium phonon distribution $n^{eq}(T)$ , (3) the matrix
$\mat{B}$ describing incoherent scattering and (4) the transmission
coefficients
$\mat{{\mathcal{T}}}$ for ballistic transport.

The limitations of Eq.~\ref{eq:conductancelandauer} for the
conductance have been pointed out by Simons\cite{simons74_jpc7_4048}
and further extended by others
\cite{chen98_prb57_14958,landry09_prb80_165304,merabia12_prb86_94303}:
Eq~\ref{eq:conductancelandauer} predicts a finite thermal resistance
even for an interface between identical materials. This paradox is
resolved by including on both sides the deviations
$n_\lambda-n^{eq}_\lambda(T)$ from thermal equilibrium occupations as
obtained from the Phonon-Boltzmann equation
Eq.~\ref{eq:phononboltzmann}.

%=====================================================================
\section{Equations of atomic motion and phonons}
\label{sec:eqm}
%=====================================================================
The theory presented so far uses the phonon density $n_\lambda$ as
the basic quantity. Here we establish contact of the phonon distribution
with the dynamics of individual atoms. 

Let us start from the Newton's equations of motion for the atoms in a
crystal. We denote the atomic coordinates by $R_{i,\vec{t}}$, where
$i=1,\ldots,3N$ and $N$ is the number of atoms in a unit cell.  With
$\vec{t}$ we denote the lattice vectors
$\vec{t}_{i,j,k}=\vec{T}_1i+\vec{T}_2 j+\vec{T}_3 k$ given by the three
primitive translation vectors $\vec{T}_1,\vec{T}_2,\vec{T}_3$ of the
crystal and integer $i,j,k$.

The potential energy is given by the energy surface $V(\vec{R})$,
which depends on all atomic positions in the crystal.
Expanding the total energy in a Taylor expansion about a minimum up to
second order provides the force constants
$c_{i,\vec{t},j,\vec{t'}}=\frac{\partial^2 V} {\partial
  R_{i,\vec{t}}\partial R_{j,\vec{t'}}}$.  The force constants define
the phonon modes, whereas higher-order terms in this Taylor expansion
describe phonon-phonon scattering.

The Lagrangian for the displacements $u_{i,\vec{t}}$ from the
equilibrium positions is
\begin{eqnarray}
\mathcal{L}=\frac{1}{2}\sum_{i,\vec{t}}m_i\dot{u}^2_{i,\vec{t}}
-\frac{1}{2}
\sum_{i,j,\vec{t},\vec{t'}}
u_{i,\vec{t}}
c_{i,\vec{t},j,\vec{t'}}
u_{j,\vec{t'}}
\;,
\label{eq:generallagrangianharmonic}
\end{eqnarray}
where $m_i$ are the atomic masses. The sum over lattice translation
vectors $\vec{t}$ includes $\mathcal{N}$ unit cells.

The equations of motion for the displacements are obtained as
Euler-Lagrange equations
$\frac{d}{dt}\frac{\partial\mathcal{L}}{\partial\dot{u}_{i,\vec{t}}}
-\frac{\partial\mathcal{L}}{\partial{u}_{i,\vec{t}}}=0$, which yields
\begin{eqnarray}
m_i\ddot{u}_{i,\vec{t}}=-\sum_{j,\vec{t'}}
c_{i,\vec{t},j,\vec{t'}}u_{j,\vec{t'}}\;.
\label{eq:neqm}
\end{eqnarray}

The discrete translational symmetry of a crystal is reflected in the
equation
\begin{eqnarray}
c_{i,\vec{t},j,\vec{t'}}=c_{i,\vec{0},j,\vec{t'}-\vec{t}}
\end{eqnarray}
for the force constants. The eigenvalue for a translation by $\vec{t}$
is $\e{i\vec{k}\vec{t}}$, which is expressed by the wave vector
$\vec{k}$. 

The displacements can be described by the eigenmodes of the lattice
translation operator in the form
\begin{eqnarray}
u_{i,\vec{t}}(t)&=&\frac{1}{\sqrt{\mathcal{N}}}
\sum_{\vec{k},\sigma}
\frac{1}{\sqrt{m_i}}
U_{i,\sigma}(\vec{k}) 
\e{i\vec{k}\vec{t}} Q_\sigma(\vec{k},t)\;,
\nonumber\\
\label{eq:ansatzphonon}
\end{eqnarray}
where $Q_\sigma(\vec{k},t)$ are the mode amplitudes. The sum contains
$\mathcal{N}$ wave vectors. The mode amplitudes are related to each
other by the condition $Q^*(\vec{k},t)=Q(-\vec{k},t)$, enforcing real
displacements. 

The parameters $U_{i,\sigma}(\vec{k})$ for a given band index $\sigma$
and wave vector $\vec{k}$ are normalized eigenvectors of the dynamical
matrix
\begin{eqnarray}
D_{i,j}(\vec{k})=\sum_{\vec{t'}}
\frac{c_{i,\vec{0},j,\vec{t}}\;\e{i\vec{k}\vec{t}}}{\sqrt{m_im_j}}
\;.
\label{eq:dynamicalmatrix}
\end{eqnarray}
They are chosen so that
$U_{i,\sigma}(\vec{k})=U^*_{i,\sigma}(-\vec{k})$.  The squared angular
frequencies $\omega^2_\sigma(\vec{k})$ are the eigenvalues of the
dynamical matrix, that is
\begin{eqnarray}
\sum_{j=1}^{3N}D_{i,j}(\vec{k})U_{j,\sigma}(\vec{k})
=U_{i,\sigma}(\vec{k})\omega_\sigma^2(\vec{k})
\;.
\end{eqnarray}

One outcome is the dispersion relation, i.e. the angular frequency
$\omega_\sigma(\vec{k})$ as function of wave vector and the index
$\sigma$ for a particular phonon branch. It yields the phonon energy
$\epsilon_\sigma(\vec{k})=\hbar\omega_\sigma(\vec{k})$ and the group
velocity
$\vec{v}_\sigma(\vec{k})=\vec{\nabla}_k\omega_\sigma(\vec{k})$.

In terms of the mode amplitudes $Q_\sigma(\vec{k},t)$, the Lagrangian
can be expressed in terms of individual harmonic oscillators.
\begin{eqnarray}
\mathcal{L}&=&\sum_{\vec{k},\sigma}
\frac{1}{2}\biggl[
 \dot{Q}^*_{\sigma}(\vec{k})\dot{Q}_\sigma(\vec{k})
-\omega^2_\sigma(\vec{k})
Q^*_\sigma(\vec{k})Q_\sigma(\vec{k})
\biggr]
\end{eqnarray}

The mode amplitudes can be written as
\begin{eqnarray}
Q(\vec{k},t)&=&\sqrt{\frac{\hbar}{\omega_\sigma(\vec{k})}}
\nonumber\\
&\times&\Bigl(
b_\sigma(\vec{k})\e{-i\omega_\sigma(\vec{k})t}
+b^*_\sigma(-\vec{k})\e{+i\omega_\sigma(\vec{k})t}\Bigr)
\;.
\label{eq:qtob}
\end{eqnarray}
While the coefficients $Q_\sigma(\vec{k})$ describe waves travelling
in opposite directions, Eq.~\ref{eq:qtob} rearranges the modes such
that the phonon amplitudes $b_\sigma(\vec{k},t)$ describe waves that
travel along only one specific direction.

The total energy is a sum over the individual phonons
\begin{eqnarray}
E=\sum_{k,\sigma}\hbar\omega_\sigma(\vec{k}) b^*_\sigma(\vec{k})b_\sigma(\vec{k})
\label{eq:etotphononclassical}
\end{eqnarray}
so that the phonon density of a mode $\lambda=(\vec{k},\sigma)$,
\begin{eqnarray}
n_\lambda=\frac{1}{\mathcal{N}\Omega_T}b^*_\lambda b_\lambda
\;,
\label{eq:phonondensity}
\end{eqnarray} 
can be expressed by the absolute square of the phonon amplitudes
$b_\lambda$.  $\Omega_T$ is the volume of a real-space unit cell.  The
constant $n^0_\lambda=1/(\mathcal{N}\Omega_T)$ is the density per
phonon in mode $\lambda$.

Note, that Eqs.~\ref{eq:etotphononclassical} and
  \ref{eq:phonondensity} are still purely classical expressions.  In
a quantum description, the phonon amplitudes translate into creation
and annihilation operators and the energies would be shifted by the
zero-point energy. The thermal average of the squared amplitudes
$b^*_\lambda b_\lambda$ is replaced by the statistical expectation
value of the corresponding quantum expression, namely the Bose
distribution $\langle{b}^\dagger_\lambda{b}_\lambda\rangle_T
=\left[\e{\beta\epsilon_\lambda}-1\right]^{-1}$.

%=====================================================================
\section{Interfaces and complex band structure}
\label{sec:complexbandsstructure}
%=====================================================================
%=====================================================================
\subsection{Conservation of $k_{||}$ at interfaces}
%=====================================================================
If a material has an internal interface or a surface, the
translational symmetry perpendicular to the interface or surface is
broken.  Thus, modes with different wave vectors $\vec{k}_\perp$
perpendicular to the surface are no more independent. However, one can
still characterize the solutions by the wave-vector components
parallel to this plane, described by $\vec{k}_{||}$.

The fact that $\vec{k}_{||}$ is a preserved quantity, allows one to
break down the problem of a multilayer structure with parallel
interfaces into many quasi-one-dimensional problems, one for each
value of $\vec{k}_{||}$. The possibility to divide the problem into
effective one-dimensional problems is an enormous simplification and
will be exploited even when this symmetry is weakly broken.

In order to judge the relevance of the $k_{||}$ selection rule, one
should distinguish the two-dimensional surface unit cell of the
individual half-crystals from the two-dimensional unit cell of the
interface: Depending on the registry of the two half-crystals, the
interface unit cell may contain several surface unit cells of each
half-crystal.

Interesting is the point of view of one particular half crystal.  An
incoming phonon with a parallel wave vector $\vec{k}_{||}$ in the
surface reciprocal unit cell of one half crystal produces reflected
phonons not only for the same value of $\vec{k}_{||}$, but also for
any vector $\vec{k}_{||}+\vec{g}^{I}_{||,1}i+\vec{g}^{I}_{||,2}j$
shifted by a reciprocal lattice vector of the interface. A poor
fit of the two lattices implies a large
real-space interface unit cell and hence small reciprocal lattice
vectors for the interface.

As the interface becomes more complex, i.e. for larger interfacial
unit cells, the spacing of this grid of reciprocal interfacial lattice
vectors decreases. Thus, in the limit of an incommensurate interface
this grid covers all space, which renders the selection rule
meaningless.

%=====================================================================
\subsection{Complex wave vectors and complex band structure}
%=====================================================================
In an extended material, only modes with real wave vectors are of
relevance, because only these lead to finite amplitudes in an infinite
crystal. At an interface, however, we also need to consider modes that
fall off towards the interior of the
material\cite{tamm32_pzsu1_733}. Such evanescent modes either couple
to extended modes on the other side of the interface, or they form
interface states that fall off in both directions away from the
interface.

These are modes with complex perpendicular wave vector $k_\perp$. They
describe wave functions that fall off exponentially away from the
interface. Each of these evanescent modes has a partner that falls
off in the opposite direction.

\begin{figure}[h!]
\includegraphics[width=\linewidth,clip=true]{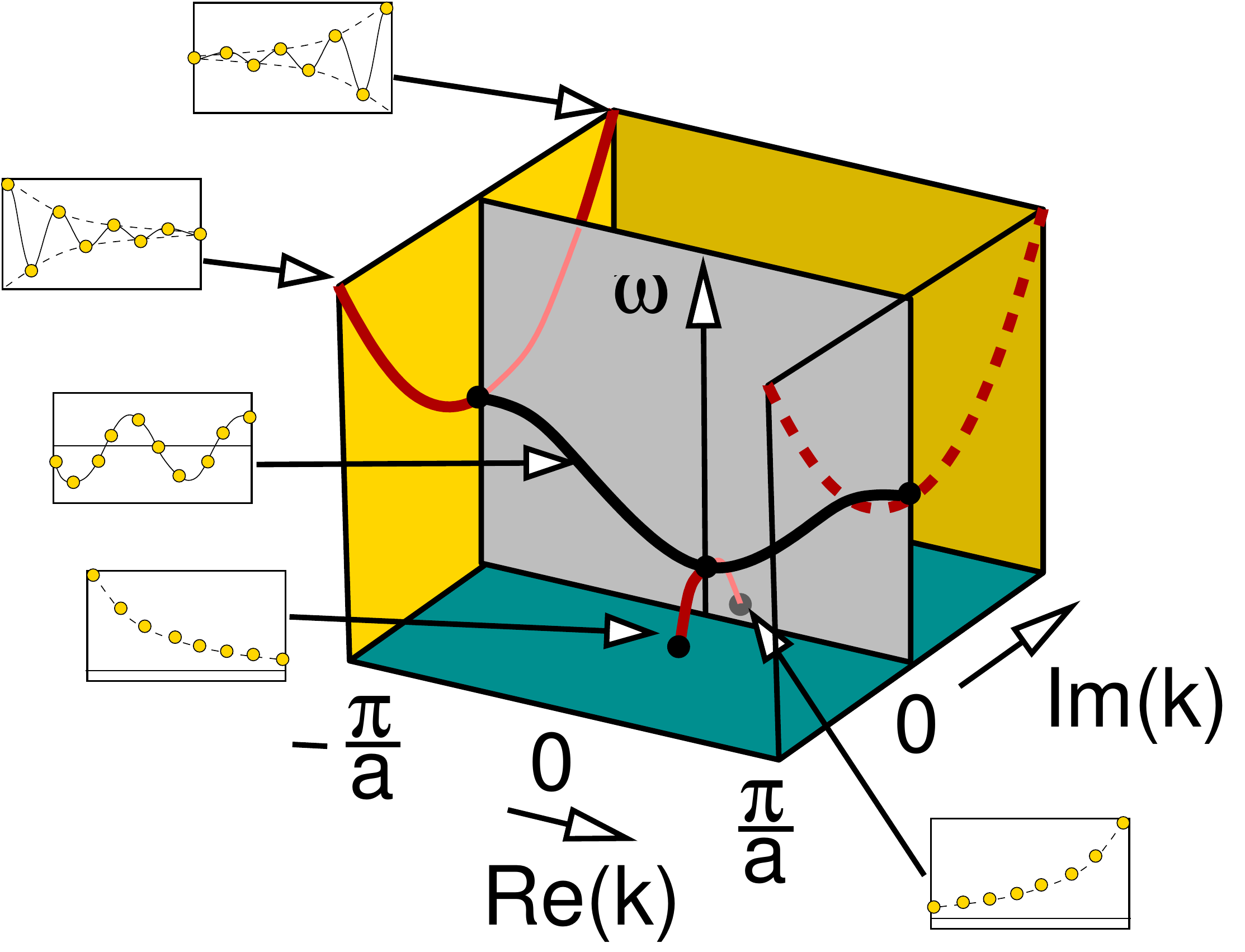}
\caption{\label{fig:dispersion}Complex band structure for the
  one-dimensional hanging linear chain described in
  section~\ref{eq:hanginglinearchain}. The black line is the real
  band structure, which describes extended solutions. The black and red
  lines together form the complex band structure. The solutions with
  complex wave vector have either an exponential decay or growth.  The
  band structure is periodic in the $\Re(k_\perp)$ direction.  The
  dashed red line is a periodic replica of the red
  line at the other Brillouin-zone boundary.}
\end{figure}

A dispersion relation $\omega_\sigma(\vec{k})$ that considers not only
the extended solutions with real wave vector, but also those with
complex wave vector is called a \textit{complex
  band structure}\cite{kohn59_pr115_809,heine63_ppsl81_300}. For a
given $k_{||}$, a complex band structure consists of lines in the
three-dimensional $(\Re(\omega)$, $\Re(k_\perp)$, $\Im(k_\perp))$
space.  An example of a complex band structure alongside with sketches
of the corresponding displacement fields is shown in
Fig.~\ref{fig:dispersion}.

%=====================================================================
\subsection{Complex eigenvalue problem}
%=====================================================================
The dynamical matrix for complex wave vectors is not necessarily
hermitian. As a result, the eigenvalues are not real-valued and
the eigenvectors are not orthogonal. Rather, the eigenvalues
$\omega^2_\sigma(\vec{k})$ are complex, and there are distinct
left-handed and right-handed eigenvectors. Nevertheless, left-handed
eigenvectors $\vec{y}^{(L)}_{\sigma}$ and right-handed eigenvectors
$\vec{y}^{(R)}_{\sigma}$ are still related by a bi-orthogonality
condition, i.e.
\begin{eqnarray}
\left(\vec{y}\,^{(L)}_{\sigma}\right)^*\vec{y}\,^{(R)}_{\sigma'}=0
\qquad\text{for $\sigma\neq\sigma'$}\;,
\end{eqnarray}
which allows one to construct one set from the other by a matrix
inversion and subsequent normalization.

Diagonalization of the dynamical matrix with a given real wavevector
$\vec{k}_{||}$ in the interface plane and a complex wave vector
$k_{\perp}$ perpendicular to the plane yields 
\begin{eqnarray}
\omega^2_\sigma(\vec{k}_{||},\Re(k_\perp),\Im(k_\perp))
\end{eqnarray}
with a complex-valued squared frequency $\omega^2_\sigma$. In the
following we will use the term complex band structure also for this
object.

%=====================================================================
\section{Model systems}
\label{sec:modelsystes}
%=====================================================================
In order to demonstrate the principles and the working of the method
we have chosen three model systems, namely the hanging linear chain,
the cubic spring model and a one-dimensional multi-band model
that avoids high symmetries.
%
%=====================================================================
\subsection{Hanging linear chain}
\label{eq:hanginglinearchain}
%=====================================================================
A minimal model for a phonon dispersion relation is the hanging
linear chain. The hanging linear chain is a one-dimensional chain of
pendula connected by springs.

In contrast to the conventional linear chain, the real band structure
of the hanging linear chain has a minimum frequency above zero. This
qualitative feature is typical for the quasi-onedimensional problems
resulting from interfaces with non-zero wave vector in the plane
$k_{||}$.  While the acoustic phonon branch of a three-dimensional
phonon band structure reaches down to zero at $k_{||}=0$, the
band structure for $k_{||}\neq0$ has a non-zero minimum frequency.

We start from a Lagrangian
\begin{eqnarray}
\mathcal{L}
=\sum_{j=-\infty}^\infty
\biggl[
\frac{1}{2}m\dot{u}_j^2
-\frac{1}{2}c\Bigl(u_{j+1}-u_j\Bigr)^2
-\frac{1}{2}bu_j^2\biggr]\;,
\end{eqnarray}
where the first term describes the kinetic energy and the remaining
terms are, with opposite sign, the potential energies
stored in the springs and pendula.

The equation of motion is
\begin{eqnarray}
m\ddot{u}_j = c \Bigl(u_{j+1}-2u_j+u_{j-1}\Bigr)- b u_j\;,
\label{eq:eqmhlchain}
\end{eqnarray}
where $j$ is the index of the bead along the chain. With a lattice
constant $a$, the equilibrium position of the $j$-th bead is at $ja$.
The term $-b u_j$ is the linearized restoring force of the
pendulum.

The dynamical matrix depends on the complex wave vector $k$, which
corresponds to $k_\perp$ in a three-dimensional problem.  The
eigenvalues of the dynamical matrix for this system are
\begin{eqnarray}
\omega^2(k)=\frac{b}{m} +\frac{2c}{m}\left(1-\cos(ka)\right)\;,
\end{eqnarray}
where $a$ is the lattice constant, respectively the spacing of the
beads.  This complex band structure is shown schematically in
Fig.~\ref{fig:dispersion}.

The complex band structure is given by
\begin{eqnarray}
k=\pm
\begin{cases}
\frac{i}{a}\arccosh(F(\omega))
&\text{for $\omega<\sqrt{\frac{b}{m}}$}
\\
\frac{1}{a}
\arccos(F(\omega))&\text{for $\sqrt{\frac{b}{m}}$}
<\omega<\sqrt{\frac{b+4c}{m}}
\\
\frac{\pi}{a}
+\frac{i}{a}\arccosh(-F(\omega))
&\text{for $\omega>\sqrt{\frac{b+4c}{m}}$}
\end{cases}
\nonumber\\
\end{eqnarray}
with $F(\omega)=\frac{b+2c-m\omega^2}{2c}$.  Because the dynamical
matrix is a scalar, its eigenvector $U_{j,\sigma}(k)$ used in
Eq.~\ref{eq:ansatzphonon} is a number, namely one.

Let us highlight two
observations\cite{kohn59_pr115_809,heine63_ppsl81_300} that we will
return to in the following.
\begin{enumerate}
\item the number of bands are the same for all frequencies. (In this
  case two.);
\item bands emerge into the complex plane at the extrema of the real
  band structure.
\end{enumerate}

%=====================================================================
\subsection{Cubic spring model}
\label{eq:3dsquaremodel}
%=====================================================================
In order to go beyond one-dimensional models, we use the cubic spring
model. This model consists of atoms arranged on a simple cubic lattice
with nearest-neighbor bond stretch and bond bend terms.  The high
symmetry of the model simplifies calculations.

The lattice constants are $\vec{T}_1=(a,0,0)$, $\vec{T}_2=(0,a,0)$ and
$\vec{T}_3=(0,0,a)$.

The Lagrangian is given by Eq.~\ref{eq:generallagrangianharmonic} for
one atom per unit cell, i.e. $i,j\in\{1,2,3\}$, and the force constant
matrix
\begin{eqnarray}
c_{i,\vec{0},j,\vec{t}}=
\delta_{i,j}\biggl[
c\Bigl(\delta_{\vec{t},\vec{T}_i}
-2\delta_{\vec{t},\vec{0}}
+\delta_{\vec{t},-\vec{T}_i}\Bigr)
\nonumber\\
+\bar{c}\sum_{k\neq i}
\Bigl(\delta_{\vec{t},\vec{T}_k}
-2\delta_{\vec{t},\vec{0}}
+\delta_{\vec{t},-\vec{T}_k}\Bigr)\Bigr]\;.
\end{eqnarray}
The spring constant $c$ describes the force constant for bond
stretching and determines the longitudinal speed of sound.  The spring
constant $\bar{c}$ on the other hand describes the force constant for
bond bending and determines the transversal speed of sound. 

Below, we use this model to calculate the spectral distribution of the
conductance for the current from material $A$ to material $B$. The
parameters are $m^A=m^B=1$, $a^A=a^B=1$, $c^A=0.125$,
$\bar{c}^A=0.05$, $c^B=0.075$, $\bar{c}=0.02$, where the super script
indicates the respective material at the interface. The spring
constant across the interface are the average spring
constants of both materials.

The eigenvalues of the dynamical matrix are
\begin{eqnarray}
\omega^2_\sigma(\vec{k})&=&
\frac{2c}{m}\Bigl(1-\cos(k_\sigma a)\Bigr)
\nonumber\\
&+&\sum_{\sigma'; \sigma'\neq\sigma}\frac{2\bar{c}}{m}\Bigl(1-\cos(k_\sigma a)\Bigr)
\;,
\end{eqnarray}
where $\sigma$ is one of the three polarization directions. The
eigenvectors of the dynamical matrix are
$U_{i,\sigma}(\vec{k})=\delta_{i,\sigma}$

For an interface parallel to one of the cubic lattice planes, the
band structure along for $k_\perp$ can be mapped for each $k_{||}$ onto
the hanging linear chain described above. The parameters are given in
table~\ref{tab:fromcubtohanginglinear}.

\begin{table}[h!]
\begin{center}
\begin{tabular}{|c|c|c|}
\hline
polarization & $c$ & $b$\\
\hline
(1,0,0) & $\bar{c}$ & $2c\cos(k_xa)+2\bar{c}\Bigl(1-\cos(k_ya)\Bigr)$ \\
(0,1,0) & $\bar{c}$ & $2\bar{c}\cos(k_xa)+2c\Bigl(1-\cos(k_ya)\Bigr)$ \\
(0,0,1) &      $c$  & $2\bar{c}\cos(k_xa)+2\bar{c}\Bigl(1-\cos(k_ya)\Bigr)$ \\
\hline
\end{tabular}
\end{center}
\caption{\label{tab:fromcubtohanginglinear}Parameters $(b,c)$ for the
  hanging linear chain model described in
  section~\ref{eq:hanginglinearchain}, equivalent to the cubic spring
  model with parameters $c,\bar{c}$ for a given interface wave vector
  $k_{||}=(k_x,k_y)$. The masses $m$ and lattice constants $a$ of both
  systems are identical.}
\end{table}

%=====================================================================
\subsection{One-dimensional  3-band model}
\label{sec:threebandmodel}
%=====================================================================
A third model system is used for testing the numerical methods for
complex band structure.  This model has been devised to contain
several qualitative features present in real systems such as (1)
several extrema in real bands (2) loops in the complex plane that
connect two bands at different real wave vector and (3) more than the
minimum number (2) of vertical bands.  Thus, this model lacks a number
of symmetry-dictated properties of the hanging linear chain.

\begin{figure}[h!]
\begin{center}
\includegraphics[width=\linewidth]{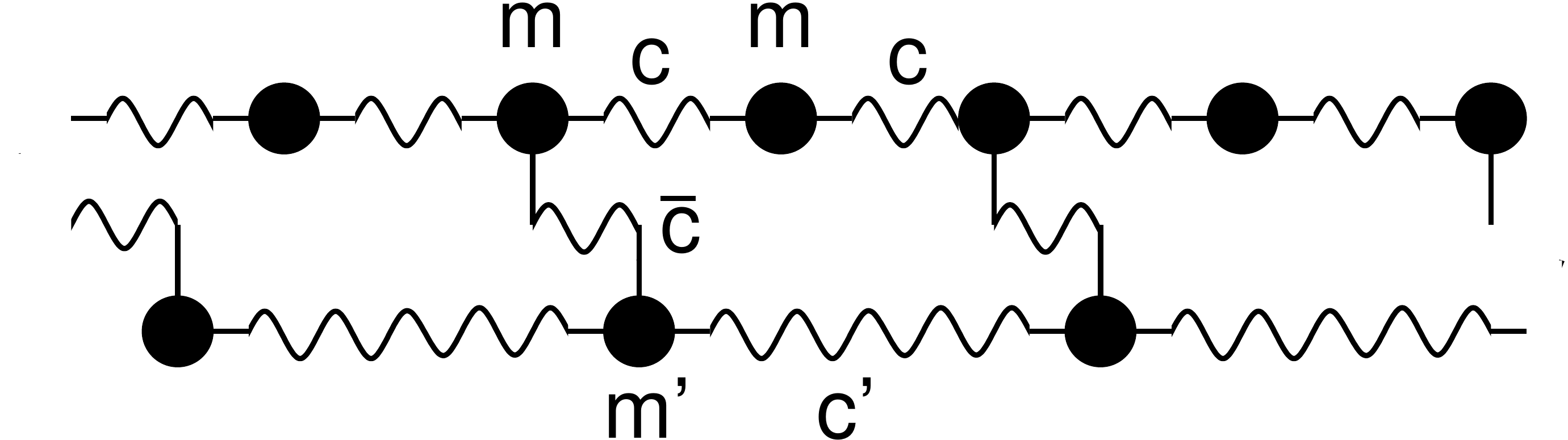}
\end{center}
\caption{\label{fig:3bandmodel} Sketch of the 3-band model described
  in section~\ref{sec:threebandmodel}.}
\end{figure}

The model consists of two coupled linear chains: One of the chains
is described by the Lagrangian
\begin{eqnarray}
\mathcal{L}_1=\sum_j
&\biggl(&\frac{1}{2}
\sum_{i=1}^2 m\dot{u}_{i,j}^2 -\frac{1}{2}c (u_{1,j}-u_{2,j})^2
\nonumber\\
&+&\frac{1}{2}c (u_{2,j}-u_{1,j+1})^2\biggr)\;.
\label{eq:3bandl1}
\end{eqnarray}
It has two atoms in the unit cell and produces one acoustic and one
 optical branch.  The other chain, described by the Lagrangian
\begin{eqnarray}
\mathcal{L}_2&=&\sum_j\biggl(\frac{1}{2} m'\dot{u}_{3,j}^2 
-\frac{1}{2}c' (u_{3,j}-u_{3,j+1})^2\biggr)\;
\label{eq:3bandl2}
\end{eqnarray}
 has only one atom per cell and thus contributes only an acoustic
 branch.  A spring couples the second chain to every second 
 bead of the first chain.
\begin{eqnarray}
\mathcal{L}_{3}&=&\sum_j\biggl(-\frac{1}{2}\bar{c} (u_{1,j}-u_{3,j})^2\Biggr)
\label{eq:3bandl3}
\end{eqnarray}
The coupling avoids the crossings and results in three distinct bands
in the real band structure.

The total Lagrangian
$\mathcal{L}=\mathcal{L}_1+\mathcal{L}_2+\mathcal{L}_3$ is the sum of
the three terms.  All parameters of the model, $m,m',c,c',\bar{c}$,
are set equal to one.

The complex band structure of this system has been calculated with the
methods described below and is shown in Fig.~\ref{fig:cband}.

%=====================================================================
\section{Numerical determination of the complex band structure}
\label{sec:numericscomplexbands}
%=====================================================================
The numerical methods for evaluating the complex band structure of
phonons are closely related to those for electrons.  The standard
method to determine the complex band structure is the eigenvalue
method developed by Chang and
Shulman\cite{chang82_prb25_605,chang82_prb25_3975}. It has been used
to describe the complex band structure of electrons and has later been
extended to phonons\cite{yip84_prb30_7037}. The eigenvalue method is
an efficient method to determine the complex bands energy by
energy. In contrast, the method described below follows the complex
bands in complex $k_\perp$-plane. Thus our method seems suitable for
the description of phonon transport, which is, unlike electrons, not
limited to a small energy region near a Fermi surface.

%=====================================================================
\subsection{Triangulation}
\begin{figure}[htb]
\begin{center}
\includegraphics[width=\linewidth]{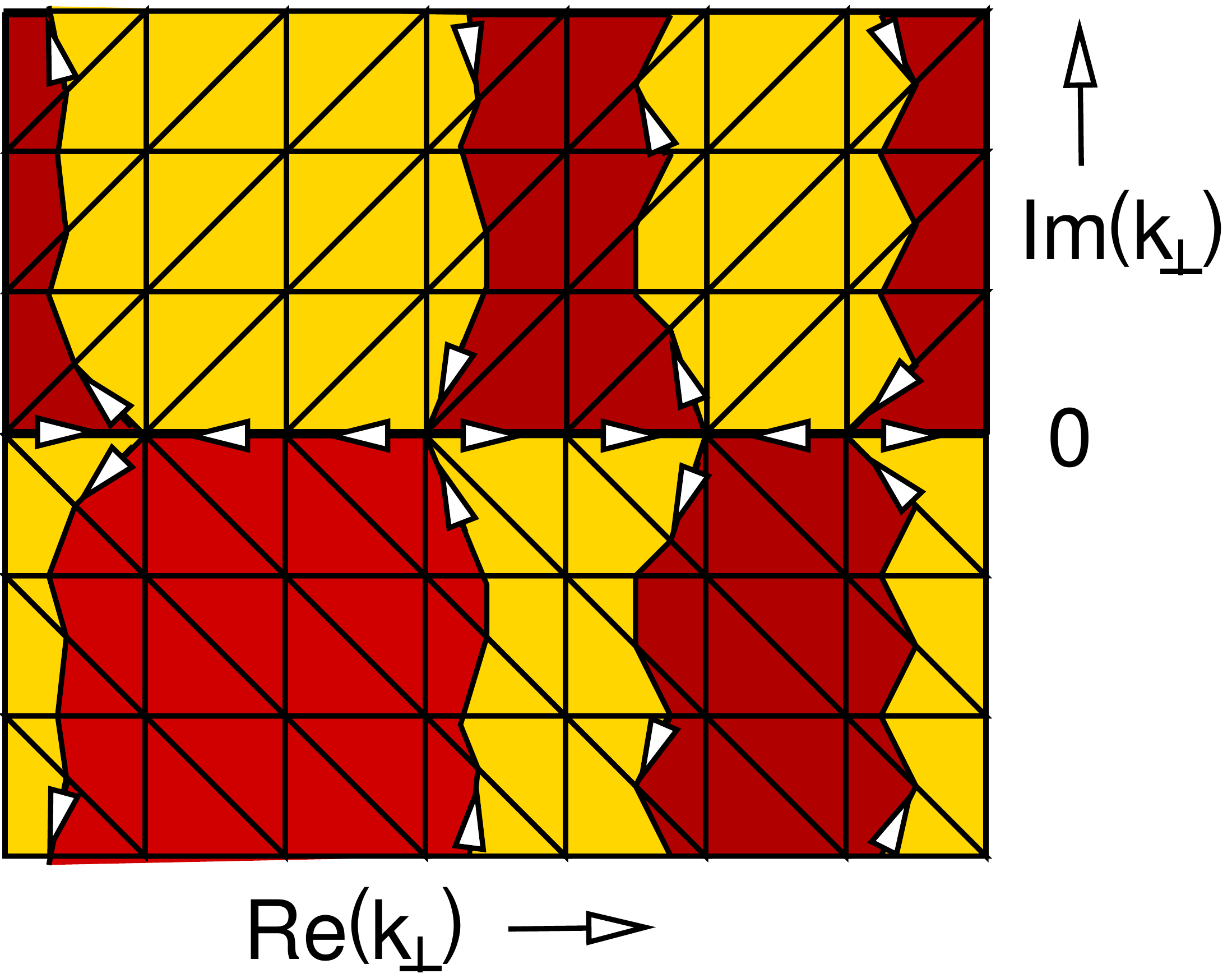}
\end{center}
\caption{\label{fig:triangulation} Triangulation of the complex
  $k_\perp$ plane. The regions with different color indicate the
  different signs of $\Im(\omega^2_\sigma)$ for a specific band
  index $\sigma$. The real part $\Re(\omega^2_\sigma)$ of the squared
  frequencies increases counterclockwise around the dark regions
  with positive imaginary part $\Im(\omega^2_\sigma)$ as indicated by
  the arrows.}
\end{figure}

%=====================================================================
In order to address the problem numerically, we define a discrete grid
of points in the complex $k_\perp$-plane.  The region in between the
grid points is triangulated, so that the triangles fill the complex
$k_\perp$ plane up to a minimum negative and a maximum positive
imaginary value. As shown in Fig.~\ref{fig:triangulation}, the
triangles are arranged with a mirror symmetry in the real axis in
order to exploit the corresponding symmetry of the complex band
structure.

Inside each triangle, $\omega^2_\sigma$ is linearized.  Thus, the
information at the grid points is sufficient to determine the
triangulated $\omega^2_\sigma$ surfaces completely.

This linearization is the main numerical approximation in the
method. It can easily be controlled by choosing a finer grid.  In the
following, the dispersion relation is replaced by its triangulated
approximation, which simplifies the topology of the complex band
structure and rigorously avoids certain crossings of complex bands.

The diagonalization of the dynamical matrix for a given grid point
$k_\perp$ is one of the most time critical steps in the calculation.
Only a small fraction of the grid points in the complex $k_\perp$
plane is visited, however, so that the diagonalization need not be
performed for all grid points. Rather, the diagonalization for a given
$k_\perp$ point is performed only when needed. Once obtained, the
information is kept for later use.

%=====================================================================
\subsection{Branch cuts and spurious bands}
%=====================================================================
Frequencies with a finite imaginary part describe solutions that
either decay or grow exponentially with time. They are of no
interest for our problem. Thus our first task is to find lines in the
complex $k_\perp$-plane with $\Im(\omega^2_\sigma)=0$.

At this point, one faces the problem that the choice of the branch
index $\sigma$ is not unique. Common library routines return
eigenvalues ordered with increasing real part. For this choice, we
find that the imaginary part of $\omega_\sigma^2$ is discontinuous and
the real part has discontinuous derivatives.  Such a jump of
$\Im(\omega_\sigma^2)$, which changes its sign, produces spurious
solutions of $\Im(\omega^2_\sigma)=0$.

The behavior described above, namely the existence of branch cuts, is
typical for avoided band crossings and characteristic of the way
different branches of the complex band structure are connected to each
other.

\begin{figure}[h!]
\begin{center}
\includegraphics[width=0.4\linewidth]{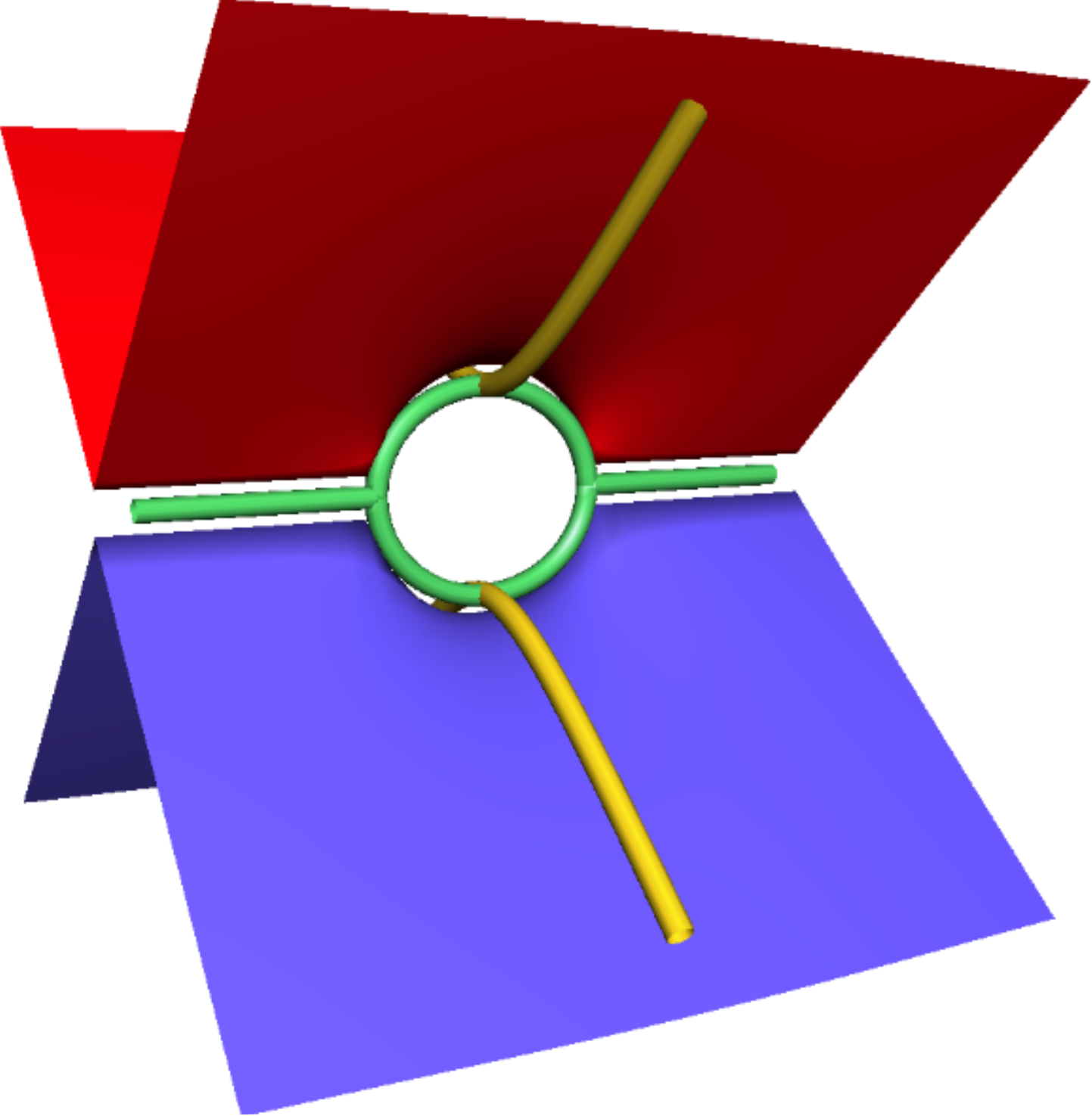}
\includegraphics[width=0.45\linewidth]{fig5b}
\end{center}
\caption{\label{fig:avcross}Real (left) and imaginary part
  (right) of the squared frequency $\omega^2=\pm\sqrt{k^2_\perp+1}$ at an
  avoided crossing. The real parts of the two branches are shifted
  vertically by a small amount to make the complex bands visible. The
  band for real $k_\perp$ is shown in yellow, while the one extending
  into the complex plane is shown in green. The bands pointing
  horizontally away from the green circle are spurious bands.}
\end{figure}

Let us therefore demonstrate the behavior for an idealized avoided
band crossing in greater detail for a minimal model of the dynamical
matrix which depends linearly on the complex $k_\perp$, namely
\begin{eqnarray}
\mat{D}(k_\perp)=\left(\begin{array}{cc} k_\perp & 1\\1 \;&
  -k_\perp\end{array}\right)
\end{eqnarray}
The characteristic equation yields $\omega^2=\pm\sqrt{k^2_\perp+1}$. The result
for real and imaginary part of $\omega^2$ in the complex plane and the
resulting complex bands are shown in Fig.~\ref{fig:avcross}.

In this example, the lines with $\Im(\omega^2_\sigma)=0$ are the two
orthogonal lines with either $\Re(k_\perp)=0$ or $\Im(k_\perp)=0$.  At
the line, where the two sheets of $\Re(\omega^2_\sigma(k_\perp))$
cross, i.e. at $\Re(k_\perp)=0$ and $|\Im(k_\perp)|>1$, the imaginary
part of $\omega^2_\sigma$ exhibits a discontinuity and each sheet of
$\Im(\omega^2_\sigma(k_\perp))$ changes its sign.  At the onset of this
discontinuity, i.e. at $k_\perp=\pm i$, are the points in the complex
$k_\perp$ plane, where the complex bands connect from one sheet to the
other. These are also the points where the branch cut, respectively
the discontinuity of the imaginary part of $\omega^2$ start to
develop, and where the spurious bands, the straight sections of the
green lines, lie.

The two $\omega^2_\sigma$ sheets are multiply connected. Trying to
treat them individually causes discontinuities either in the real part
or the imaginary part of $\omega^2$ surfaces. Where exactly these
discontinuities lie, depends on the way the frequencies are attributed
to different bands. However, the discontinuities can not be avoided
altogether when separating different bands.

Treating all bands together from the outset makes the concepts rather
complicated. A surprisingly simple solution to this problem has been
to ignore the problem at first and to divide the bands according to
increasing $\Re(\omega_\sigma^2(k_\perp))$ into different
sheets. Then, the complex bands are extracted for each sheet
individually. This introduces on the one hand spurious bands, which,
on the other hand, are easily removed later on, when the band from
different sheets are connected.

\begin{figure}[h!]
\begin{center}
\includegraphics[width=0.4\linewidth]{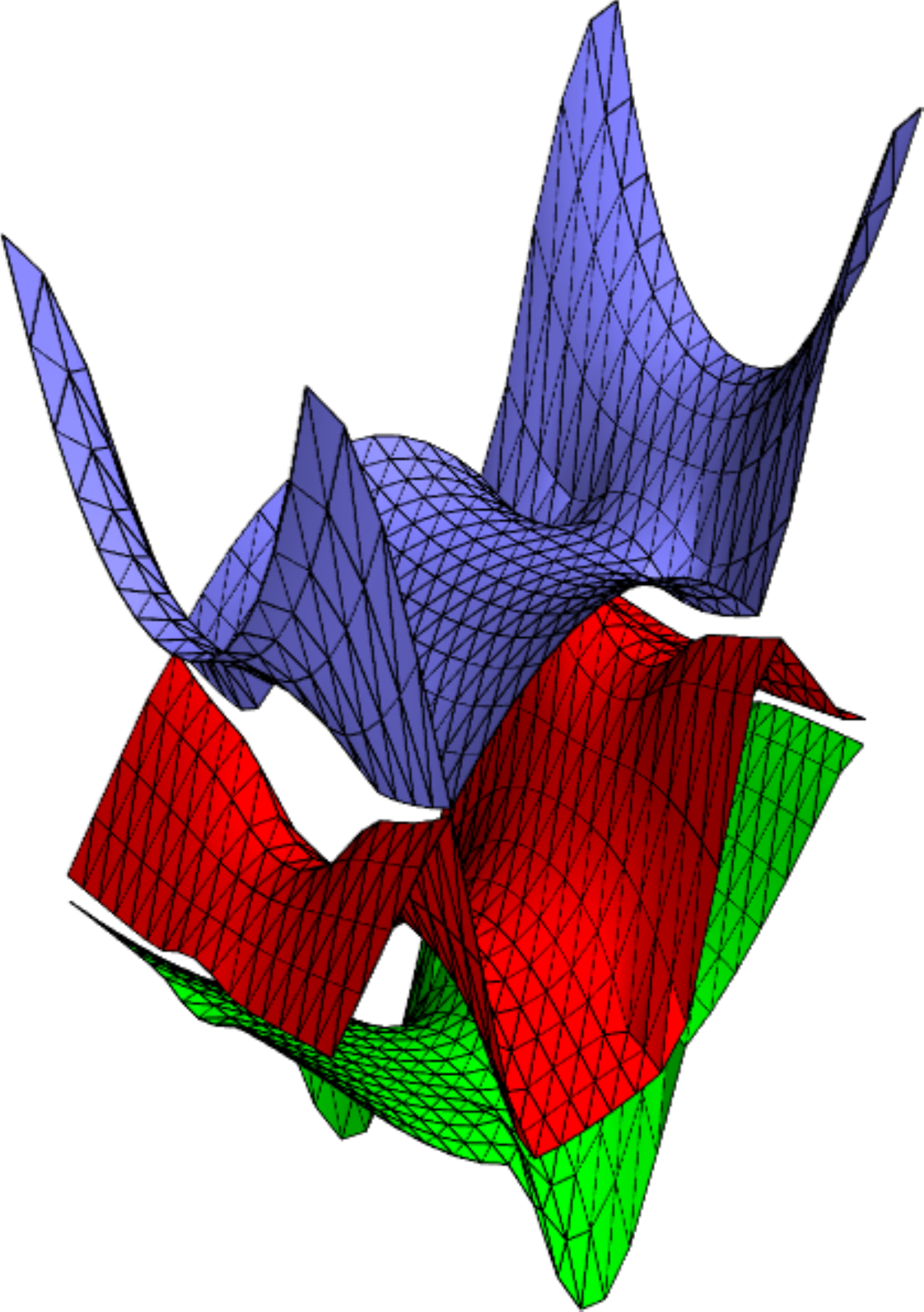}
\includegraphics[width=0.4\linewidth]{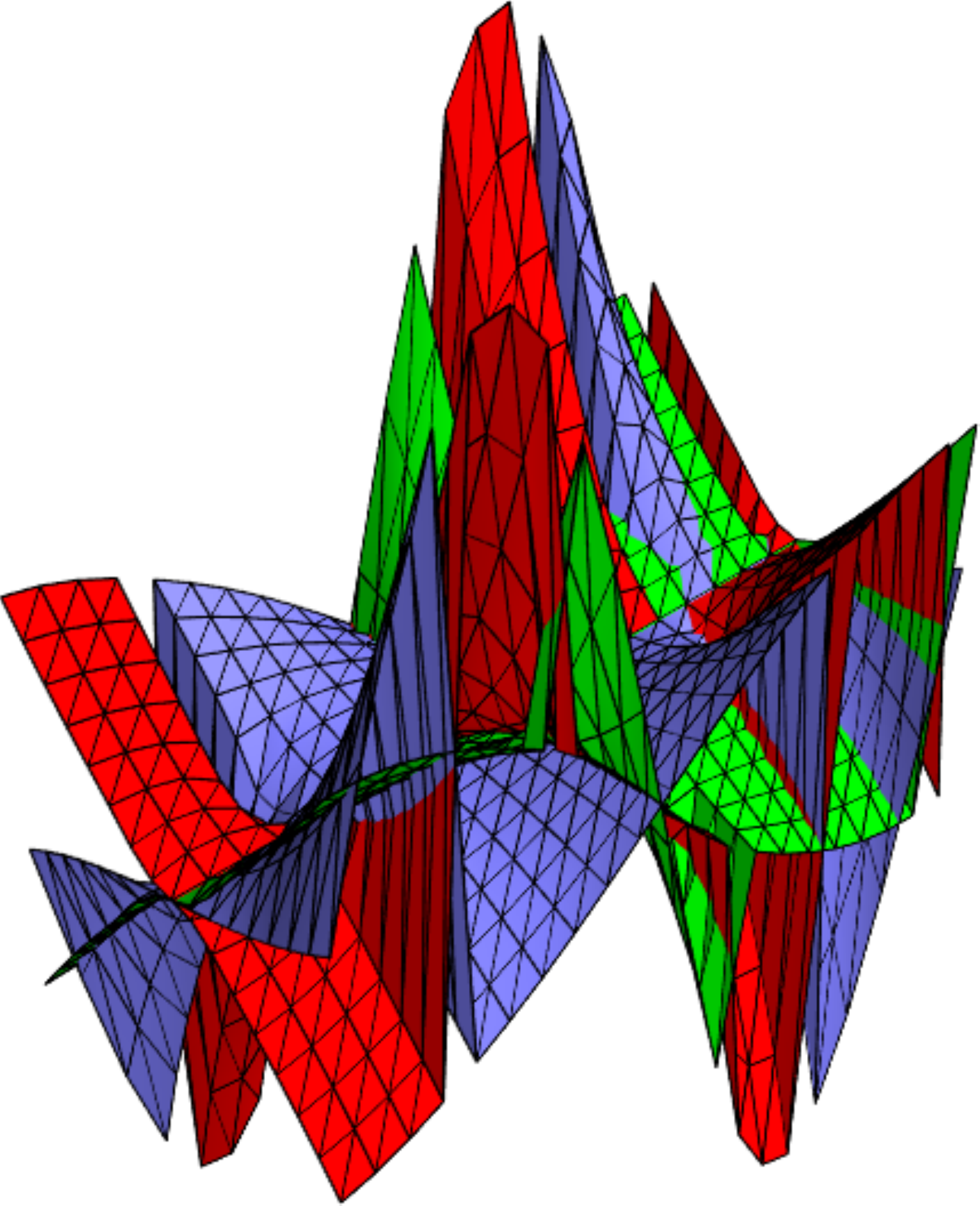}
\end{center}
\caption{\label{fig:reimomega2}Real part (left) and imaginary part
  (right) of $\omega_\sigma^2$ in the complex $k_\perp$-plane. If the
  the eigenvalues of the dynamical matrix are ordered with respect to
  increasing real part, the imaginary part is connected in a
  discontinuous manner. The resulting sign changes produce spurious
  bands that need to be removed.}
\end{figure}

%=====================================================================
\subsection{Analytic properties of the complex band structure 
and vertical bands}
%=====================================================================
The analytic properties of the complex band structure for the
one-dimensional Schr\"odinger equation has been analyzed by
Kohn\cite{kohn59_pr115_809}. Heine\cite{heine63_ppsl81_300}
generalized the results to higher dimensions. They are summarized for
example by Yip and Chang\cite{yip84_prb30_7037}.
 
The complex band structure has a mirror symmetry of the real part and
a mirror antisymmetry of the imaginary part of $\omega^2_\sigma$ at
the plane with $\Im(k_{\perp})=0$,
i.e. $\omega^2_\sigma(k_\perp)=\left(\omega^2_\sigma(k^*_\perp)\right)^*$,
where $z^*$ denotes the complex conjugate of
$z$.\cite{yip84_prb30_7037,blount62_ssp13_305}.

From the Cauchy-Riemann equations for a given band $\omega_\sigma^2$
follows that the condition $\Im[\omega^2_\sigma]=0$ determines
directions of steepest descent of
$\Re[\omega^2_\sigma(\vec{k}_\perp)]$. The path following the gradient
of $\Re[\omega^2_\sigma(\vec{k}_\perp)]$ proceeds in counterclockwise
direction around a region with $\Im[\omega^2_\sigma]>0$. The
monotonous growth of a $\Re[\omega^2_\sigma(\vec{k}_\perp)]$ while
circling around a region with positive $\Im[\omega^2_\sigma]$ excludes
disconnected regions with one sign of the imaginary part, which in
turn excludes disconnected complex bands.  These properties do not
hold on a branch cut, which violates the Cauchy-Riemann equations.

The finding above indicates that the topology of the complex band
structure can be comprehended reasonably well by exploring the regions
in the complex plane with $\Im[\omega^2]>0$. 

An important consequence of the above is that branches of the complex
band structure emerge away from the real band structure at the extrema
of of the latter \cite{heine63_ppsl81_300}.  Furthermore, the bands
emerge with increasing $\Re(\omega^2_\sigma)$ at the maxima of the
real band structure and with decreasing $\Re(\omega^2_\sigma)$ at the
minima.

We will use this property to construct \textit{vertical bands} that
grow monotonously in $\Re(\omega^2_\sigma)$ from zero to infinity.
These vertical bands are then used to extract the eigenmodes of each
material for a specific frequency in order to match them at
an interface plane.

The construction of vertical bands is facilitated by the following:
The conditions $\Im(\omega^2_\sigma)>0$ and $\Im(\omega^2_\sigma)<0$
divide the complex $k_\perp$ plane into distinct regions. The complex
band structure lives on the boundaries of these regions. 

%=====================================================================
\subsection{Algorithm for single sheets}
%=====================================================================
Let us now outline the construction of vertical bands representing the
complex band structure for a given $k_{||}$.  The goal is to obtain
sequences of $(\Re(\omega_\sigma^2),\Re(k_\perp),\Im(k_\perp))$, which
constitute the ends of the linear line segments of the complex band
structure.

We start out evaluating the real band structure, i.e. the eigenvalues
of the dynamical matrix with $\Im(k_\perp)=0$. The eigenvalues provide
bands with real-valued, positive $\omega^2_\sigma$.

The following steps are performed for each band index $\sigma$
individually:
\begin{enumerate}
\item We start at the first grid point of the real band structure and
  proceed in one direction along the real $k_\perp$-axis,
  accumulating a sequence of $k_\perp$ points and their
  $\Re(\omega^2_\sigma)$-values. When an extremum of $\omega^2_\sigma$
  is encountered, we leave the real $k_\perp$ axis and follow the line
  with $\Im(\omega^2_\sigma)$ into the complex plane.  For this
  purpose, we search the grid points next to the real $k_\perp$ axis
  for a pair of values with opposite signs of
  $\Im(\omega_\sigma^2)$. Linear interpolation between this pair
  yields one end of the line segment which is connected to the
  extremum of the real band structure on the real axis. This is the
  first line segment protruding away from the real band structure.
\item From this line segment, the branch is grown to the boundary of
  the complex $k_\perp$-grid. This construction is unique: Each line
  segment of the complex band structure enters a specific
  triangle. For this triangle, the eigenvalues of $\omega^2_\sigma$
  are known for two corners of the triangle. After diagonalizing the
  dynamical matrix for the third corner, the linear line segment with
  $\Im(\omega^2_\sigma)=0$ within this triangle is calculated
  analytically from the linearized $\omega^2_\sigma(k)$. If
  $\Im(\omega^2_\sigma)$ vanishes accidentally at a grid point we
  change this value $\Im(\omega^2_\sigma)$ artificially to an very
  small negative value. This modification guarantees that the line
  segment leads to one of the sides of the triangle and that it does
  not pass through a grid point. The
  process of growing the complex band is repeated for the next
  triangle.  The growth process for the complex band ends at the
  end-points of the grid in the complex $k_\perp$ plane.
\item The complex bands are calculated only for $\Im(k_\perp)\ge0$.
  The corresponding branch for $\Im(k_\perp)<0$ is obtained from the
  mirror symmetry
 $\omega_\sigma^2(k_\perp)=\left(\omega_\sigma^2(k_\perp^*)\right)^*$ of
    the complex band structure at the plane with $\Im(k_\perp)=0$.
    This band, with the points arranged in the opposite order, is the
    first section of the next vertical band for this band index
    $\sigma$. It leads back to the real axis, from where we continue
    to accumulate the $k_\perp$ points and $\Re(\omega^2_\sigma)$
    points until we reach the next extremum. From there we again
    proceed into the complex plane, but now to the same side of the
    $k_\perp$ grid, where this section of the complex band started
    from.
\item This procedure is repeated until we encounter the end of the
  real axis. Using periodic boundary conditions, we connect the
  unfinished band with the first band, which we started from the first
  grid point on the real $k_\perp$-axis.
\end{enumerate}

Thus we obtain a number of segments of the vertical band structure
shown in the bottom of Fig.~\ref{fig:cband}. Each segment starts at
the outer boundary of the grid in the complex $k_\perp$ plane and
proceeds inward to the real $k_\perp$ axis. Then it follows from one
extremum of the real band structure to the next, from where it turns
again back into the complex plane out to the outer boundary. As long
as the the segments of the vertical band structure stay on one side of
the complex plane, their $\Re(\omega^2)$ values either increase or
decrease monotonously.

%=====================================================================
\subsection{Connect different branches}
%=====================================================================
So far, we obtained for each band index $\sigma$ 
a number of sections of
vertical bands. These sections are now connected between different
band indices.

We find that complex bands of two band indices tend to arrive in pairs
at the boundary of our $k_\perp$-grid. As in
Fig.~\ref{fig:reimomega2}, these two lines first form two halves of a
closed line connecting maxima and minima of the real band structure.
This circle, however, is not closed but both bands lead nearly
straight and parallel away to the outer bounds of our
$k_\vert$ grid.

We first connect these pairs of lines from different sheets. We make
use of the fact that they arrive within the same triangles at the
boundary. This leads to vertical bands running over the entire
frequency range.

In the following step we remove the loop of spurious bands describing
the branch cut, which lead from the circles to the boundary of the
complex plane and back. Thus we arrive at a representation of the
complex band structure in terms of monotonously increasing vertical
bands.

%=====================================================================
\subsection{Demonstration of the method for a multi-band model}
\begin{figure}[h!]
\includegraphics[width=\linewidth,clip=true]{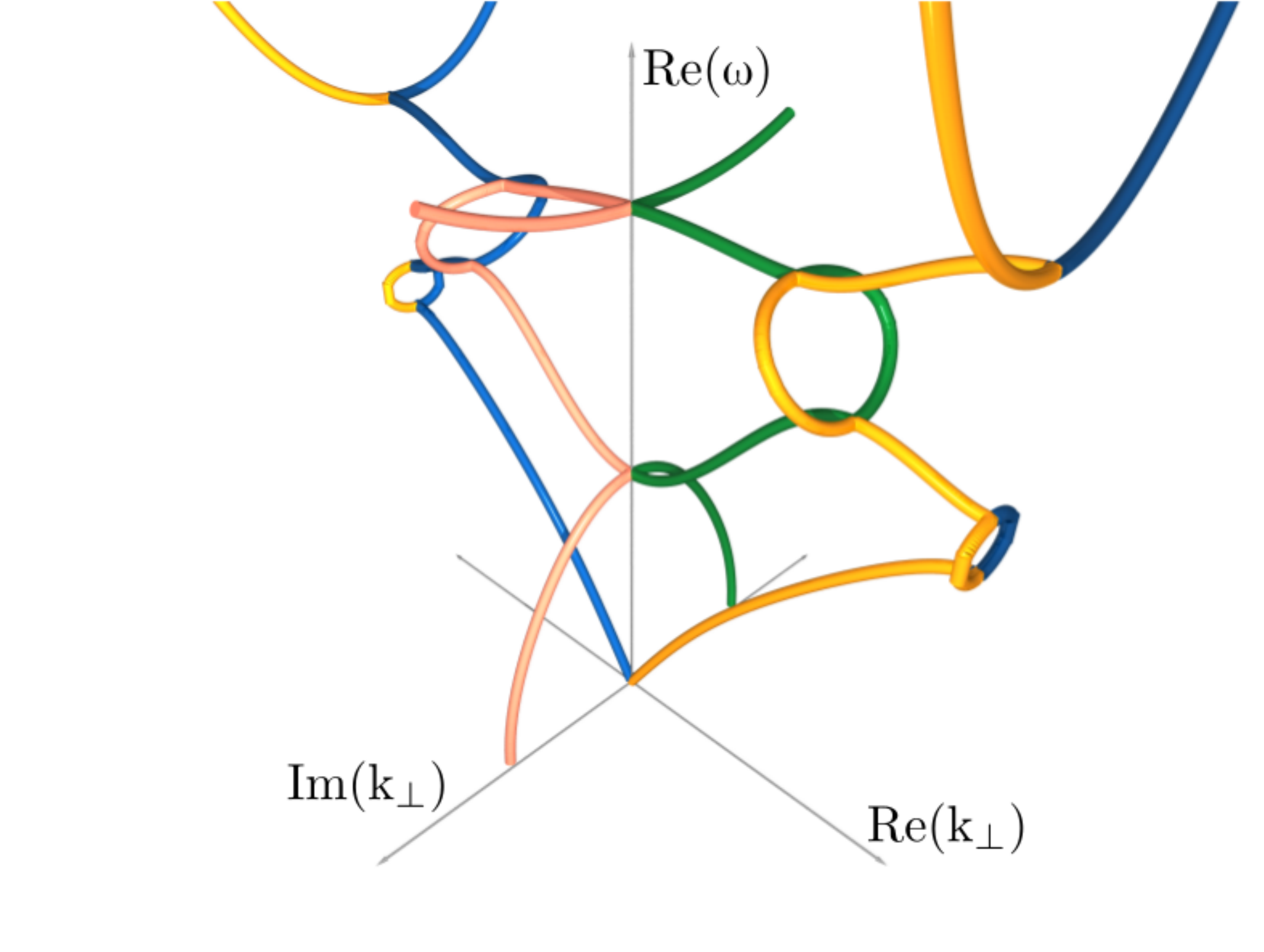}
\includegraphics[width=\linewidth,clip=true]{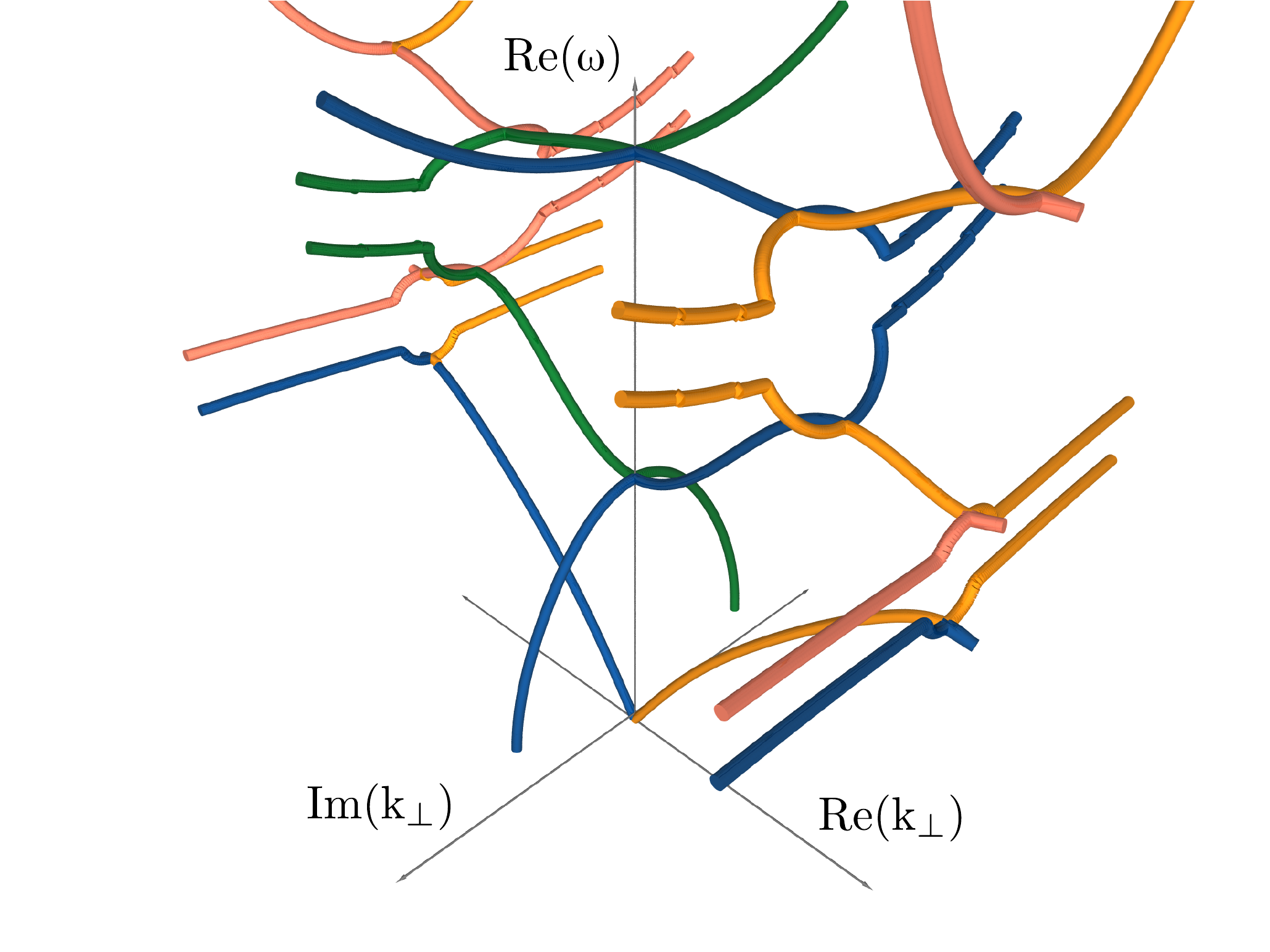}
\caption{\label{fig:cband}Top: complex band structure of the
  three-band model described in
  section~\ref{sec:threebandmodel}. Different colors indicate
  different vertical bands of the complex band structure.  Bottom:
  Same as above but before connecting different branches $\sigma$.
  Bands from different branches are shifted vertically away from each
  other for better visibility. Visible are the spurious bands at the
  branch cuts.}
\end{figure}

%=====================================================================
The methodology described above has been applied to the three-band
model described in section~\ref{sec:threebandmodel}. The resulting
complex band structure is shown in Fig.~\ref{fig:cband}.

%=====================================================================
\section{Beam matching}
\label{sec:beammatching}
%=====================================================================
Once the complex band structures for the two materials at the
interface have been evaluated, the partial solutions from the two
materials need to be matched across the interface.

This problem is addressed for a given frequency and $\vec{k}_{||}$.
The vertical bands of the complex band structure provides convenient
access to the complex wave vectors $k_{\perp,\alpha}(\omega)$ and the
corresponding modes for a given $\omega$. In the following, we will
suppress the dependence on $\omega$.  We will furthermore use the wave
vectors
$\vec{k}_\alpha(\omega)=\vec{k}_{||}+k_{\perp,\alpha}(\omega)\vec{e}$,
where $\vec{e}$ is the interface normal, and where the index $\alpha$
labels the vertical bands in the complex band structure of a given
material.

We start with the equations of motion Eq.~\ref{eq:neqm} for the
displacements at the interface, which are Fourier transformed in time.
\begin{eqnarray}
\sum_{j,\vec{t'}}
\biggl(D_{i,\vec{t},j,\vec{t'}}-\omega^2\delta_{i,j}\delta_{\vec{t},\vec{t'}}\biggr) 
u_{j,\vec{t'}}(\omega)=0
\label{eq:eqminterface}
\end{eqnarray}
For atoms near the interface, the dynamical matrix differs from that
in the bulk, because some atoms have changed when forming the
interface. Some atoms may be displaced, some have been removed by
cutting away one half crystal, and others have been added by attaching
the half crystal of the other material.

We insert the ansatz 
\begin{eqnarray}
u^{A}_{i,\vec{t}}=
\sum_\alpha\frac{1}{\sqrt{m_i}}
\bar{U}^{A}_{i,\alpha}\e{i\vec{k}^{A}_\alpha
(\vec{t}-\vec{z}_I)} q^{A}_\alpha
\label{eq:ansatzbeammatcha}
\\
u^{B}_{i,\vec{t}}=
\sum_\beta\frac{1}{\sqrt{m_i}}
\bar{U}^{B}_{i,\beta}\e{i\vec{k}^{B}_\beta(\vec{t}-\vec{z}_I)} 
q^{B}_\beta
\label{eq:ansatzbeammatchb}
\end{eqnarray}
for the displacements of the materials $A$ and $B$ into
Eq.~\ref{eq:eqminterface}, where $\vec{k}^{A}_\alpha$ is a complex
wave vector of material $A$ for the specified frequency $\omega$. The
index $\alpha$ identifies the vertical band from which this $k$-point
is obtained. With $\bar{U}^A_{j,\alpha}$ we denote the
  right-handed eigenvectors of the dynamical matrix
  $D_{i,j}^A(\vec{k}^{A}_\alpha)$ at the specified frequency $\omega$.
The definition for $\vec{k}^{B}_\beta$ is analogous for material $B$.
The position of the interface plane is $z_I$.

To simplify the equations of motion Eq.~\ref{eq:eqminterface} of an atom,
we subtract the equation of motion of the respective bulk
materials--with the displacements Eq.~\ref{eq:ansatzbeammatcha} or
Eq.~\ref{eq:ansatzbeammatchb} of the respective material. The bulk
equation can be subtracted because the ansatz satisfies the bulk
equations exactly.

The resulting equations of motion for the displacements
$u_{i,\vec{t}}$ of an atom in material $A$ at the interface to
material $B$ have the form
\begin{eqnarray}
&&\sum_{\beta\in B}\Bigl(\sum_{(j,\vec{t'})\in B'}
D^{AB}_{i,\vec{t},j,\vec{t'}}\e{i\vec{k}_\beta(\vec{t'}-z_I)}
\bar{U}^B_{j,\beta}\Bigr)q^B_\beta
\nonumber\\
&-&\sum_{\alpha\in A}\Bigl(\sum_{j,\vec{t'}\in A'}
D^{A}_{i,\vec{t},j,\vec{t'}}\e{i\vec{k}_\alpha(\vec{t'}-z_I)}
\bar{U}^A_{j,\alpha}\Bigr)q^A_\alpha=0
\label{eq:eqmdiff}
\end{eqnarray}
The sum $(j,\vec{t'})\in B'$ describes the displacements in the
material $B$ on the opposite side of the interface . The sum
$(j,\vec{t'})\in A'$ runs over all displacements of atoms that have
been removed while forming the interface.  The sum $\beta\in B$ refers
to all vertical bands in the complex band structure of material $B$,
that is all complex $k_\perp$-values at the chosen frequency, and the
sum $\alpha\in A$ runs over all vertical bands of material $A$.
Equations analogous to Eq.~\ref{eq:eqmdiff} are set up for atoms in
material $B$ at the interface.

These equations are trivially obeyed for displacements $u_{i,\vec{t}}$
for which the matrix elements $D_{i,\vec{t},j,\vec{t'}}$ of the
  dynamical matrix are unaffected by the interface.

The equations of motion for the displacements with a changed
environment, on the other hand, contribute conditions that link the
coefficients $q^A_\alpha$ and $q^{B}_\beta$ of the bulk modes of the
two materials.

For the sake of simplicity, we have so far ignored atoms at the
interface that can be attributed to none of the two bulk
materials. The equation of motion for such an atom contributes three
unknowns, the amplitudes $u^I_{\gamma}$ of the interface atom, and
three equations, that link the new coordinates to the coefficients
$q^A_\alpha,q^B_\beta$ of the modes in the bulk materials. The
resulting equations are of the same form as Eq.~\ref{eq:eqmdiff}, and
they are treated analogously.

The resulting set of equations Eq.~\ref{eq:eqmdiff}, from the atoms of
the atoms of materials $A$ and $B$ at the interface and the interface
atoms, has the form of a set of homogeneous linear equations.
\begin{eqnarray}
\sum_j M_{i,j} q_{j,n}=0
\label{eq:singvalprob}
\end{eqnarray}
$\mat{M}$ is a rectangular, but not square, complex matrix. Its
left index runs over all equations of motion that are affected by the
interface. The right index of $\mat{M}$ runs over all mode amplitudes
$q^{A}_\alpha$ and $q^B_\beta$ of both materials and the amplitudes
$u^I_\gamma$ of the interface atoms. The index $n$ runs over all
independent modes of the interface system.

This problem is equivalent to searching the eigenvectors $\vec{q}_n$ of
a matrix $\mat{M}$ with vanishing eigenvalue. It is solved with
singular-value decomposition, which directly delivers the desired
vectors as the null-space of the matrix $\mat{M}$.  This null space is
spanned by all vectors, that are orthogonal to the rows of the matrix
$M$.

Each vector of the null space provides one set of parameters
\begin{eqnarray}
\vec{q}_j=\left(
\ldots, q^{A}_{\alpha,j},\ldots, u^I_{\gamma,j},\ldots, q^{B}_{\beta,j},\ldots
\right)
\label{eq:nullvector}
\end{eqnarray}
where $u^I_{\gamma}$ are the displacements of those interface atoms that
differ from the bulk materials on either side.

%===================================================
\subsection{Transfer matrix}
%===================================================
The general solution of Eq.~\ref{eq:singvalprob} is a superposition
$\vec{q}=\sum_j\vec{q}_j c_j$ with arbitrary parameters $c_j$.

One is often interested in the coefficient transfer matrix
$X_{\lambda,\lambda'}$, which provides the coefficients of the modes
in one material as linear combination of the mode amplitudes of the
other material.  The transfer matrix can be obtained from the null
vectors Eq.~\ref{eq:nullvector} after multiplication with the inverse
matrix of null vectors of the incoming channel.
\begin{eqnarray}
X^{B\leftarrow{A}}_{\beta,\alpha}=\sum_j q^{B}_{\beta,j}
\left(\mat{q}^A\right)^{-1}_{j,\alpha}
\end{eqnarray}

The solutions may be of the following types
\begin{enumerate}
\item modes that propagate in both materials. They can directly
  transmit energy through both materials and they contribute to the
  thermal current.
\item modes that are evanescent in one material but that contain
  impinging and reflected waves in the other. They do not directly
  contribute to thermal transport.
\item interface modes, that are evanescent in both directions. These
  modes exist only near the the interface and they fall off in both
  directions away from the interface. They describe phonons moving in
  the interface plane.
\end{enumerate}
These are ideal cases, that are present in their pure form only for
simple materials. Usually, the solutions of the interface system
combine features of all these special cases.

Modes such as evanescent waves and interface states that do not reach
another interface do not contribute to the transmission
coefficient. Nevertheless, phonon scattering at impurities or other
phonons not considered in the transfer matrix can redistribute their
energy into propagating modes, so that they nevertheless contribute
indirectly to the thermal current.

%=====================================================================
\subsection{Beam matching for hanging linear chains}
%=====================================================================
Let us demonstrate the workings of the beam matching for a simple model
system of two hanging linear chains with different band widths.

In material $A$, the parameters for the model are $m_A,c_A,b_A$, for
material $B$,  they are $m_B,c_B,b_B$. Across the
interface connecting the two materials, there is a new spring with
force constant $c_C$. We place the interface such that beads from
$j=-\infty$ to $j=0$ belong to material $A$, beads from $j=1$ to
$j=+\infty$ belong to material $B$.

The ansatz for the bulk-like atoms are
\begin{eqnarray}
u_j&=&
\begin{cases}
u_j^A&\text{for $j\le0$}
\\
u_j^B&\text{for $j\ge1$}
\end{cases}
\end{eqnarray}
where $u_j^A$ and $u_j^B$ are the displacements of the modes of the
bulk materials at the positions $j a$.
\begin{eqnarray}
u_j^A&=&\sum_{\alpha=1,2}\e{ik^A_\alpha(aj-z_I)}q^A_\alpha
\nonumber\\
u_j^B&=&\sum_{\beta=1,2}\e{ik^B_\beta(aj-z_I)}q^B_\beta
\label{eq:ansatzbulkmodelsystem}
\end{eqnarray}
$a$ is the lattice spacing and $z_I$ is the position of
the interface. We place the interface in the middle between the two
lattice planes, so that its value is $z_I=\frac{1}{2}a$.

There are two equations of motion that are not satisfied by the ansatz
for the bulk materials, namely the ones for $u_0$ and $u_1$.
\begin{eqnarray}
-m_A \omega^2 u_{0}=-c_A(u_{0}-u_{-1})+c_C(u_{1}-u_{0})-b_Au_0
\nonumber\\
-m_B \omega^2 u_{1}=-c_C(u_{1}-u_{0})+c_B(u_{2}-u_{1})-b_Bu_1
\nonumber\\
\end{eqnarray}

By subtracting the corresponding bulk equations, we obtain
two non-trivial equations 
\begin{eqnarray}
0&=&c_C(u^B_{1}-u^A_{0})-c_A(u^A_{1}-u^A_{0})
\nonumber\\
0&=&-c_C(u^B_{1}-u^A_{0})+c_B(u^B_{1}-u^B_{0})
\end{eqnarray}
that correspond to Eq.~\ref{eq:eqmdiff} after inserting the ansatz
Eq.~\ref{eq:ansatzbulkmodelsystem} for the displacements,.

We arrive at two homogeneous linear equations for four unknowns.  The
coefficients of these equations form the matrix $\mat{M}$ defined in
Eq.~\ref{eq:singvalprob}.  The two independent solutions spanning the
null space of $\mat{M}$ provide each a parameter set
\begin{eqnarray}
\vec{q}_j=\left(
\begin{array}{cccc}
q^{A}_{1,j};&q^{A}_{2,j};&q^{B}_{1,j};&q^{B}_{2,j}
\end{array}\right)
\end{eqnarray}

The superposition of the two vectors gives the general solution for
the interface system
\begin{eqnarray}
\left(\begin{array}{c}q^A_1\\q^A_2\\q^B_1\\q^B_2\end{array}\right)
&=&
\left(\begin{array}{cc}
q^A_{1,1}&q^A_{1,2}\\
q^A_{2,1} &q^A_{2,2}\\
q^B_{1,1}&q^B_{1,2}\\
q^B_{2,1} &q^B_{2,2}
\end{array}\right)
\left(\begin{array}{cc}
q^A_{1,1}&q^A_{1,2}\\
q^A_{2,1} &q^A_{2,2}
\end{array}\right)^{-1}
\left(\begin{array}{c}c_1\\c_2\end{array}\right)
\end{eqnarray}
By choosing $c_1=q^A_{1}$ and $c_2=q^A_2$ as initial condition, we can
now determine the solution at the interface and in material $B$.

The transfer matrix $\mat{X}$ translates the amplitudes $q^A_\alpha$
in material $A$ into those, $q^B_\beta$, of material $B$ via
\begin{eqnarray}
\left(\begin{array}{c}q^B_1\\q^B_2\end{array}\right)
&=&
\mat{X}^{A\leftarrow B}
\left(\begin{array}{c}q^A_1\\q^A_2\end{array}\right)
\end{eqnarray}
where the transfer matrix is
\begin{eqnarray}
\mat{X}^{A\leftarrow B}
&=&
\left(\begin{array}{cc}
q^B_{1,1}&q^B_{1,2}\\
q^B_{2,1} &q^B_{2,2}
\end{array}\right)
\left(\begin{array}{cc}
q^A_{1,1}&q^A_{1,2}\\
q^A_{2,1} &q^A_{2,2}
\end{array}\right)^{-1}
\end{eqnarray}

For the sake of completeness we list the result
\begin{eqnarray}
\mat{X}^{A\leftarrow{B}}
=\left(\begin{array}{cc}
\frac{M_{4,1}M_{1,2}-M_{1,1}M_{4,2}}{M_{3,1}M_{4,2}-M_{4,1}M_{4,3}}
&
\frac{M_{2,1}M_{4,2}-M_{4,1}M_{2,2}}{M_{4,1}M_{3,2}-M_{3,1}M_{4,2}}
\\
\frac{M_{1,1}M_{3,2}-M_{3,1}M_{1,2}}{M_{3,1}M_{4,2}-M_{4,1}M_{4,3}}
&
\frac{M_{2,1}M_{4,2}-M_{4,1}M_{2,2}}{M_{4,1}M_{3,2}-M_{3,1}M_{4,2}}
\end{array}\right)
\end{eqnarray}
with
\begin{eqnarray}
M_{1,\alpha}&=&-c_A\e{\frac{ik_\alpha^A a}{2}}+(c_A-c_C)\e{\frac{-ik_\alpha^A a}{2}}
\nonumber\\
M_{1,\beta+2}&=&c_C\e{\frac{ik_\beta^B a}{2}}
\nonumber\\
M_{2,\alpha}&=&c_C\e{-\frac{ik_\alpha^A a}{2}}
\nonumber\\
M_{2,\beta+2}&=&-c_B\e{-\frac{ik_\beta^B a}{2}}-(c_B-c_C)\e{\frac{ik_\beta^B a}{2}}
\end{eqnarray}

%=====================================================================
\subsection{Multilayer structures}
%=====================================================================
The transfer matrix of the interface connects the partial solutions on
either side of the interface. A second transfer matrix translates the
solutions from the left side of a material to the right. Thus, it
propagates the solution through the material from one interface to the
next.

This matrix is, for material $B$, given by the complex wave vectors in
material $B$.
\begin{eqnarray}
\mat{X}^B_{\beta,\beta'}(d)
=\delta_{\beta,\beta'}\e{ik^B_\beta d}
\end{eqnarray}
where $d:=z_{C\leftarrow B}-z_{B\leftarrow A}$ is the distance between
the interface planes. Some precaution to prevent over and underflows
is required when the wave vectors are complex.  A remedy for this
problem has been proposed by Ko and
Inkson\cite{ko88_semicondscitechnol3_791}.

The transfer matrix for the entire multilayer of materials
$A,B,C,\ldots,Z$ is obtained as the product of the individual
transfer matrices as
\begin{eqnarray}
q^Z_{\zeta}
=\sum_{\alpha}\Bigl(\mat{X}^Z
\mat{X}^{Z\leftarrow Y}\cdots\mat{X}^{C\leftarrow B}
\mat{X}^B\mat{X}^{B\leftarrow A}\mat{X}^A\Bigr)_{\zeta,\alpha}q^A_{\alpha}
\nonumber\\
\end{eqnarray}
The first matrix $\mat{X}^A$ propagates the coefficients relative to a
given point in contact $A$ to the $AB$ interface, which lies a
distance $d^A$ away. Similarly the matrix $\mat{X}^Z$ propagates the
coefficients from the $Y/Z$ interface to a selected point in contact
$Z$. The other coefficients are all expressed relative to their
initial interface. i.e., the $A/B$ interface for the coefficients in
material $A$.

%=====================================================================
\subsection{Complex band structure of a multilayer structure}
%=====================================================================
One may be interested in a material structure that is in itself
repeating, such as a multilayer $\ldots ABABA\ldots$ of alternating
layers of materials $A$ and $B$. We may consider a repeat unit $AB$
as the unit cell of the composite material and ask for the complex
band structure for the composite material. The information about the
complex band structure is contained in the transfer matrix of the
repeat unit.

Diagonalization of the transfer matrix
$X^{ML}:=\mat{X}^{A\leftarrow{B}}\mat{X}^{B}\mat{X}^{B\leftarrow{A}}\mat{X}^{A}$
for a given energy and $k_{||}$ yields
\begin{eqnarray}
\Bigl[X^{ML}_{\lambda',\lambda} -\e{ik^{ML}_\gamma
    d_{ML}}\delta_{\lambda',\lambda}\Bigr]
q^{A}_{\lambda,\gamma}(k^{ML}_\gamma)=0
\end{eqnarray}
where $d_{ML}$ is the thickness of the repeat unit and $k^{ML}$ is the
perpendicular wave vector for the multilayer structure.  The
eigenvalues $\e{ik^{ML}_\gamma d_{ML}}$ provide the complex wave
vectors. The real band structure can be constructed selecting modes
with real $k_\gamma$.

The eigenvectors provide the coefficients for the mode in the first
material $A$, and the mode amplitudes of the other materials can be
constructed from the corresponding transfer matrices as shown here
for material $B$.
\begin{eqnarray}
q^{B}_{\lambda,\gamma}(k^{ML})=
\sum_{\lambda'\in A}
X^{B\leftarrow{A}}_{\lambda,\lambda'}q^{A}_{\lambda',\gamma}(k^{ML})
\end{eqnarray}

%=====================================================================
\section{From the transfer matrix to the conductance}
\label{sec:fromtransfertoconduct}
%=====================================================================
%=====================================================================
\subsection{Transmission coefficient}
%=====================================================================
The transfer matrix $X_{\lambda,\lambda'}$ of the complete structure
links the amplitudes of the incoming phonon mode $\lambda$ of one
contact with those of the transmitted phonons in mode $\lambda'$.

The mode-resolved transmission coefficient links the outgoing phonon
flux in contact $B$ to the incoming phonon fluxes in contact $A$.  The
phonon modes $\lambda$ are labelled as incoming $A_i,B_i$ or outgoing
$A_o,B_o$ according to their velocity $v_\lambda$ perpendicular to the
interface. If contact $B$ is to the right of contact $A$, the
right-moving phonons in $A$ are incoming and the right-moving contacts
in $B$ are outgoing.

\begin{eqnarray}
\left(\begin{array}{c}\vec{q}^{B_o}\\\vec{q}^{B_i}\end{array}\right)
=\left(\begin{array}{cc}
\mat{X}^{B_o\leftarrow A_i}&\mat{X}^{B_o\leftarrow A_o}\\
\mat{X}^{B_i\leftarrow A_i}&\mat{X}^{B_i\leftarrow A_o}
\end{array}\right)
\left(\begin{array}{c}\vec{q}^{A_i}\\\vec{q}^{A_o}\end{array}\right)
\end{eqnarray}
The requirement that there are no incoming currents from contact $B$,
i.e.  $\vec{q}^{B_i}=0$, relates the outgoing phonon currents in $B$
to the incoming currents in contact $A$. 
\begin{eqnarray}
\vec{q}^{B_o}&=&\mat{S}^{B\leftarrow{A}}\vec{q}^{A_i}
\end{eqnarray}
The proportionality constant is the corresponding subblock
$\mat{S}^{B\leftarrow{A}}$ of the scattering matrix.
\begin{eqnarray}
\mat{S}^{B\leftarrow{A}}&=&
\mat{X}^{B_o\leftarrow A_i}
\nonumber\\
&-&
\mat{X}^{B_o\leftarrow A_o}
\left(\mat{X}^{B_i\leftarrow A_o}\right)^{-1}\mat{X}^{B_i\leftarrow A_i}
\end{eqnarray}

The currents are related to the squared amplitudes and the
phonon velocities. Thus, the transmission coefficient is 
\begin{eqnarray}
\mathcal{T}_{\lambda\in A/in,\lambda'\in B/out}=
\frac{1}{n^0_\lambda\vec{e}\vec{v}_\lambda}
|S^{B\leftarrow{A}}_{\lambda,\lambda'}|^2
\Bigl(n^0_{\lambda'}\vec{e'}\vec{v}_{\lambda'}\Bigr)
\end{eqnarray}
Non-propagating modes, those with a finite imaginary part of the wave
vector, are excluded from the transmission coefficient, because
the net current of all non-propagating modes vanishes.

%=====================================================================
\subsection{Interface conductance between two cubic spring models}
\begin{figure}[th]
\begin{center}
\includegraphics[width=\linewidth,clip=true]{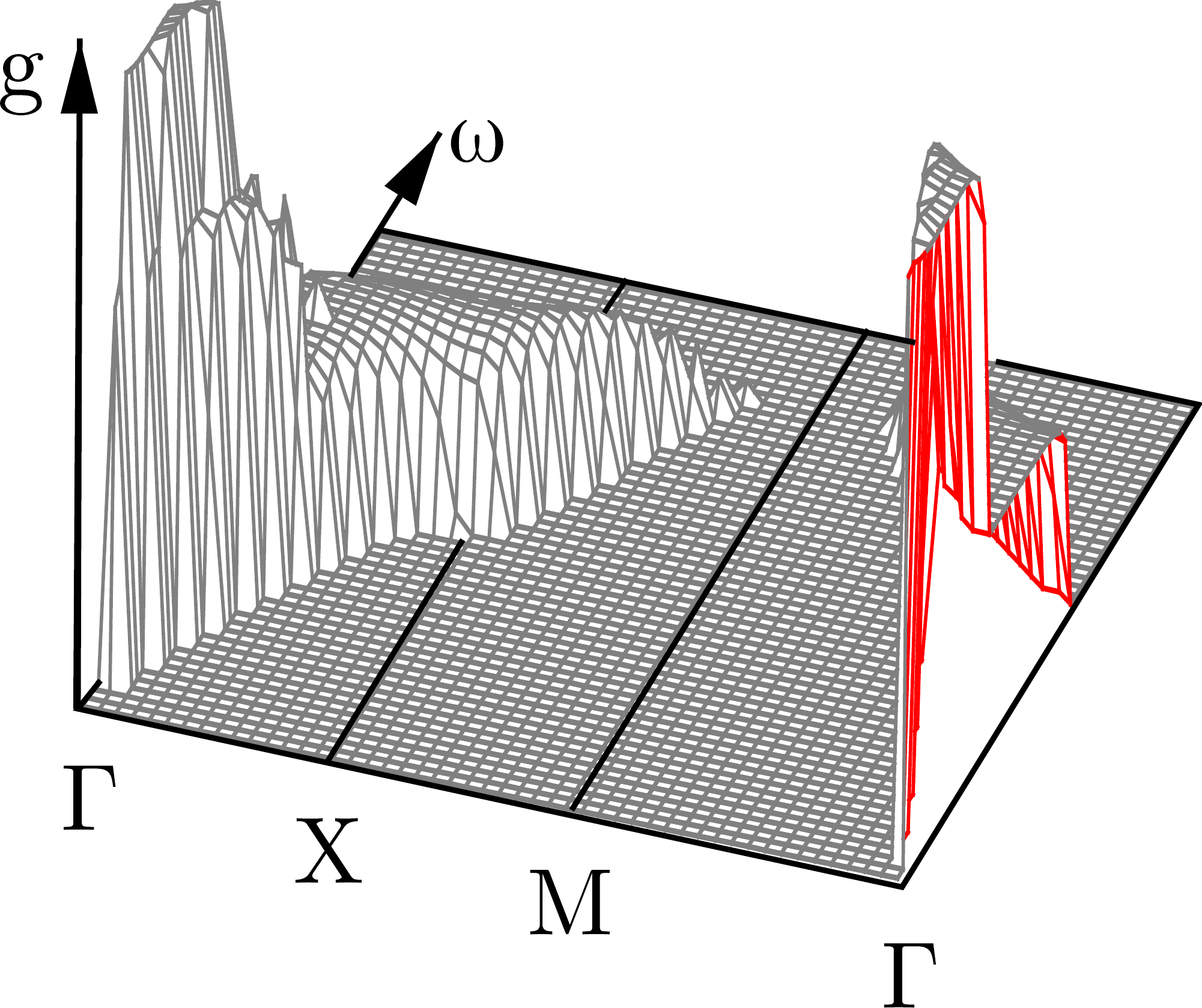}
\\[2mm]
\includegraphics[width=0.8\linewidth,clip=true]{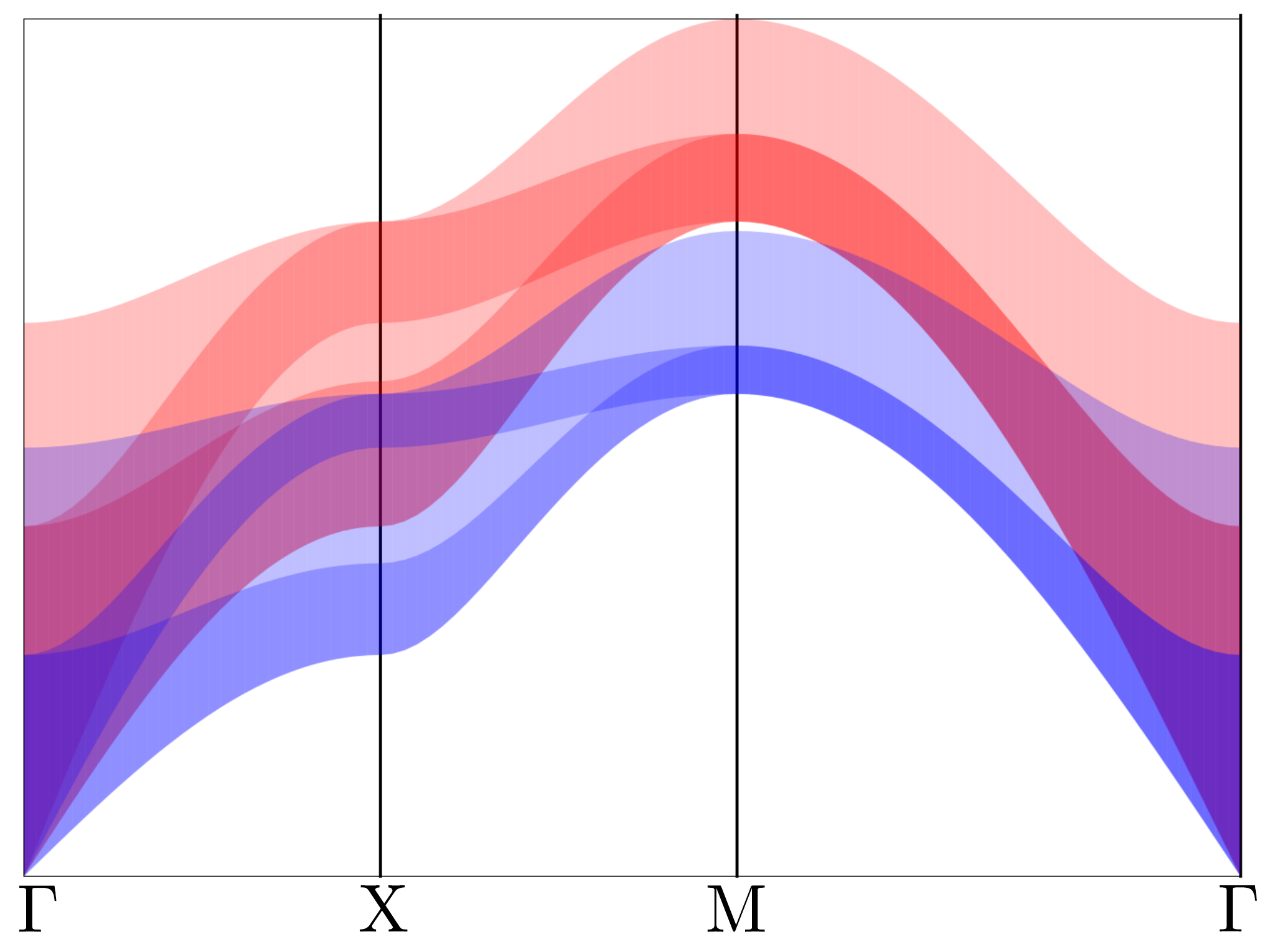}
\end{center}
\caption{\label{fig:kapitza}Top: wave-vector and energy resolved
  conductance $g(\hbar\omega,\vec{k}_{\vert\vert})$ of an interface
  between two cubic spring model systems in the high-temperature
  limit. At lower temperatures, the contribution of high-energy
  phonons is suppressed by heat capacity $C_\lambda(T)$ shown in
  Fig.~\ref{fig:heatcapacity}. Bottom: two-dimensional band structures
  $\omega_\sigma(\vec{k}_{\vert\vert})$ for hot material (red) and the
  cold material $B$ (blue). Darker regions indicate the overlap of
  several bands with different polarizations.}
\end{figure}

%=====================================================================
The ballistic interface conductance as defined in
Eq.~\ref{eq:conductancelandauer} has been analyzed for the interface
between two cubic spring models described in
section~\ref{eq:3dsquaremodel}.  The two-dimensional band structures
of both materials are shown in the bottom of
Fig.~\ref{fig:kapitza}. The parameters have been chosen to visualize
the effects of a hard-soft interface. 

In order to investigate the conductance, we introduce its density
$g(\hbar\omega,\vec{k}_{\vert\vert})$, which gives the conductance as
\begin{eqnarray}
G=\hbar\int d\omega \int d^2k_{||}\; 
\frac{1}{k_B}C(\hbar\omega,T)g(\hbar\omega,\vec{k}_{\vert\vert})
\end{eqnarray}
where $C(\hbar\omega,T)=\hbar\omega
\frac{\partial}{\partial{T}}\Bigl(\e{\beta\hbar\omega}-1\Bigr)^{-1}$
is the contribution of a mode with energy $\epsilon$ to the heat
capacity and $\beta=1/(k_BT)$. The function differs only by a constant
factor from the one shown in Fig.~\ref{fig:heatcapacity}.  In the
high-temperature limit, the contribution of a phonon mode to the heat
capacity is a constant.

This density is given by
\begin{eqnarray}
g(\hbar\omega,\vec{k}_{||})
&=&A_A
\sum_{\lambda\in{A},\lambda'\in{B}}k_Bn^0_\lambda\cdot
(\vec{e}\vec{v}_\lambda)\theta(\vec{e}\vec{v}_\lambda)
\mathcal{T}^{B\leftarrow{A}}_{\lambda,\lambda'}
\nonumber\\
&&\times\delta(\hbar\omega-\epsilon_\lambda)
\delta(\vec{k}_{||}-\vec{k}_{||,\lambda})
\end{eqnarray}
Note, that the sum over phonons in contact $A$ translates into a
one-dimensional phonon density of states of contact $A$.

Only the channels $(\omega,\vec{k}_{||})$ in the overlap of the
two-dimensional band structures of both materials can transmit heat.
This is reflected in the density $g(k_{||},\omega)$ shown in
Fig.~\ref{fig:kapitza}. It is evident that already this effect cuts
off the contribution from the high-frequency part of the
spectrum. This mechanism, which is temperature independent, supports
the notion that the acoustic modes dominate heat transport.  The
density $g(k_{||},\omega)$ has a nearly step-like behavior indicating
the number of polarizations contributing to the thermal current.

%=====================================================================
\section{Summary}
%=====================================================================
We presented a method to calculate the complex band structure of
phonons. The strength of this method is its simplicity, because it
rests on the diagonalization of the  dynamical
matrix as the basic element.  The method is economical
because it explores only those regions in the complex plane, that are
are visited by the complex band structure. An exception in the current
implementation is that branch cuts are still fully explored, a step
that can be avoided.  

We presented the beam matching of phonon modes at interfaces and
multilayers. We demonstrated how it can be used to explore the thermal
transport of phonons on the basis of individual phonons, that is,
resolved with respect to energy and in-plane wave vectors
$\vec{k}_{||}$.

It has been shown how the three-dimensional real band structure of a
multilayer can be constructed from the transfer matrices of the
components and their interfaces.

The usual approach for
mode-matching\cite{chang82_prb25_605,chang82_prb25_3975,yip84_prb30_7037}
or the calculation of Greens
functions\cite{mingo03_prb68_245406,
  sadasivam14_annrevheattransfer17_89,wang08_epjb62_381,zhang07_numerheattransferb51_333}
works selectively at a given frequency at one time. This is very
convenient for fermions, for which the transport is restricted to a
small energy window at the Fermi level. This advantage is lost in the
case of phonons, for which the entire frequency spectrum
contributes. Diagonalizing matrices as in the present approach
provides the states for the entire spectrum at given wave vector in
one step. Instead of scanning over energies for each $k_{||}$, the
present method of calculating the complex band structure scans, for
each $k_{||}$, over lines in the complex $k_\perp$ plane.

The present approach differs from slab-based methods
\cite{chang82_prb25_605,chang82_prb25_3975,yip84_prb30_7037,mingo03_prb68_245406,sadasivam14_annrevheattransfer17_89,wang08_epjb62_381,zhang07_numerheattransferb51_333,ong15_prb91_174302}
which divide the material into slabs and then match the slab solutions
at each energy. The matrix dimension of the problem is determined by
the size of the slab, and the order of the matrix problem is related
to the range of the interaction in terms of slabs. The present
approach seems to be more economical, because the full translation
symmetry of the bulk materials is exploited for the calculation of the
bulk complex band structure, while the frequency-dependent matching
step is done only once for each interface between different materials.

%=====================================================================
% acknowledgements
%=====================================================================
\begin{acknowledgements}
This work has been supported by the Deutsche Forschungsgemeinschaft
through SPP 1386 and project B03 of SFB1073. The authors gratefully
acknowledge useful discussions with C. Jooss and C. Volkert
(G\"ottingen University).
\end{acknowledgements}

\appendix
\providecommand{\jr}[1]{#1}
\providecommand{\BBB}[1]{#1}
\let\textsc\BBB

\end{document}